\documentclass[twocolumn,english]{revtex4-1}
\usepackage[T1]{fontenc}
\usepackage[latin9]{inputenc}
\usepackage{verbatim}
\usepackage{amsmath}
\usepackage{amssymb}
\usepackage{graphicx}

\makeatletter

\providecommand{\tabularnewline}{\\}

\makeatother

\usepackage{babel}
\begin{document}

\title{Banana and pizza-slice-shaped mesogens give a new constrained ferromagnet universality class}

\author{Xiuqi Ma}

\affiliation{Department of Applied Mathematics and Theoretical Physics, 
Centre for Mathematical Sciences, University of Cambridge, Wilberforce Road,
Cambridge CB3 0WA, United Kingdom.}

\author{Elsen Tjhung}

\affiliation{Department of Applied Mathematics and Theoretical Physics, 
Centre for Mathematical Sciences, University of Cambridge, Wilberforce Road,
Cambridge CB3 0WA, United Kingdom.}
\email{et405@cam.ac.uk}

\begin{abstract}
It has been known that at high density, the local orientation of banana-shaped molecules shows a spontaneously bent state,
giving rise to interesting liquid-crystalline phases such as splay-bend and twist-bend.
This spontaneous bend can be modelled theoretically by allowing the bend elastic constant in the Frank elastic energy to become negative.
Here we extend this idea to polar banana and pizza-slice-shaped molecules which can also splay spontaneously.
By allowing both splay and bend elastic constants to be negative we discovered two additional new liquid crystalline phases.
In particular, using renormalization group technique, we showed that the phase transition belongs to a new constrained ferromagnet universality class.
\end{abstract}
\maketitle

Liquid crystals are usually made up of rod-shaped and head-tail symmetric molecules. 
At high enough density (or low enough temperature), the molecules tend to align in the same direction; 
this is the nematic phase~\cite{deGennes}. 
In nematics, we denote the local average orientation of the molecules with a headless unit vector, called the director field $\hat{\boldsymbol{n}}(\boldsymbol{r})$,
such that in some mesoscopic volume at $\boldsymbol{r}$ the molecules tend to align parallel or anti-parallel to $\hat{\boldsymbol{n}}(\boldsymbol{r})$.

\begin{figure}[h]
\begin{centering}
\includegraphics[width=0.8\columnwidth]{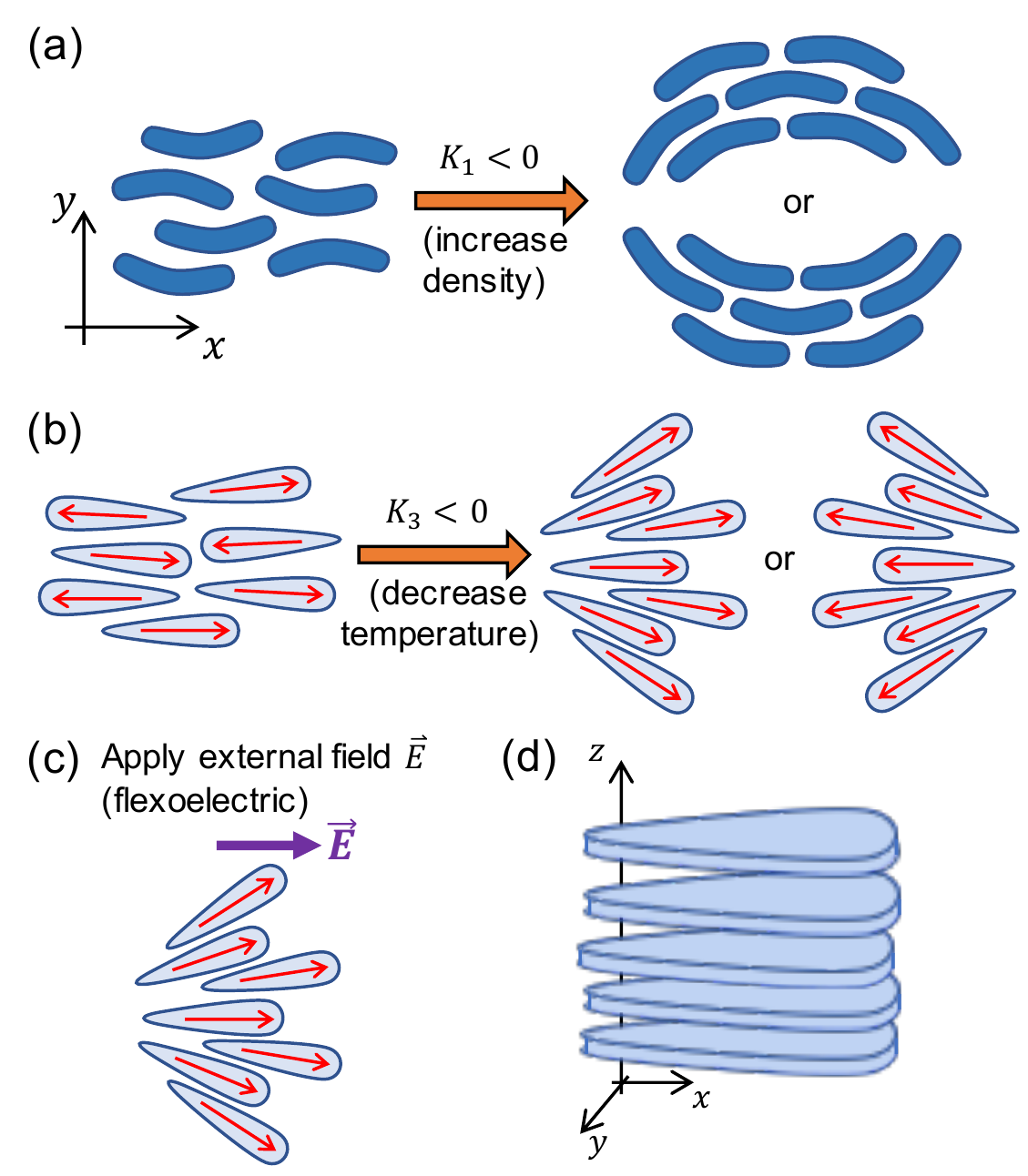}
\par\end{centering}
\caption{
(a) Banana-shaped molecules can spontaneously bend at high enough density (bend elastic constant $K_{3}$ becomes negative). 
(b) Similarly, polar pizza-slice-shaped molecules can spontaneously splay when we decrease the temperature (splay elastic constant $K_{1}$ becomes negative). 
(c) In flexoelectric liquid crystals, splay is induced by an external field not by spontaneous symmetry breaking. 
(d) The molecules form columnar stack in the $z$-direction and 
hence the director field can be assumed to be two-dimensional 
while the spatial dimension can be of dimension $d=2$ or $3$. 
(Red arrows indicate polarity of the molecules.) \label{fig:banana-pizza}}
\end{figure}

\begin{figure}[t]
\begin{centering}
\includegraphics[width=0.8\columnwidth]{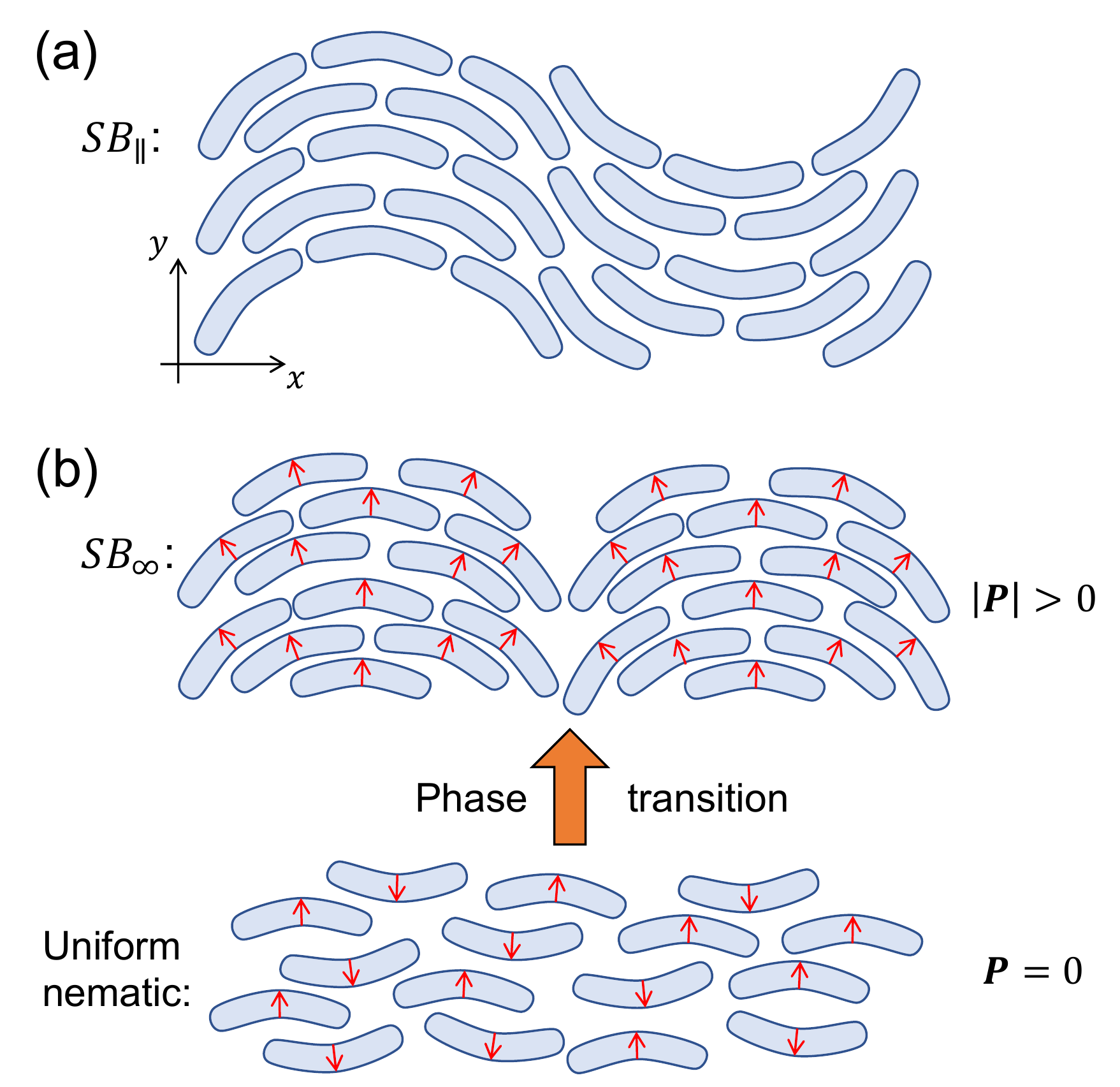}
\par\end{centering}
\caption{
(a) Apolar, head-tail symmetric banana molecules may give rise to $SB_\parallel$ phase~\cite{Dozov01}.
(b) On the other hand, polar banana molecules may give rise to $SB_\infty$ phase, which is completely different from $SB_\parallel$ phase.
(Red arrows indicate polarity of the molecules and $\mathbf{P}$ indicates total polarization.) \label{fig:polar-banana}}
\end{figure}

Now what happens if the molecules are not straight? 
For instance, one can imagine banana or pizza-slice-shaped molecules as shown in Fig.~\ref{fig:banana-pizza}.
What kind of liquid crystalline phases do they form?
The case of \emph{apolar} banana-shaped molecules has been widely studied in literature~\cite{Dozov01,Chen13,Borshch13,Wang15,Tavarone15,Memmer02,Dawood16}.
In particular, at high density, banana-shaped molecules can spontaneously bend locally~\cite{Dozov01} (see Fig.~\ref{fig:banana-pizza}(a)).
In the figure, the molecules can either bend upwards or downwards with equal probability. This is called spontaneous symmetry breaking.
At large scale, and in two-dimension, one will also get a bend modulation in the direction parallel to the director field, which is the $x$-direction in the figure.
The resulting phase is called the splay-bend parallel (or $SB_\parallel$) phase~\cite{sb}, shown in Fig.~\ref{fig:polar-banana}(a). 
In this phase, mesoscopically the system is nematic but the director field $\hat{\boldsymbol{n}}(\boldsymbol{r})$ can vary slowly in space.
This new $SB_\parallel$ phase has been verified recently in Monte Carlo simulations~\cite{Tavarone15} but not yet seen experimentally.

In three-dimension, suspensions of banana-shaped molecules can spontaneously bend to create a macroscopic helical pattern, called the twist-bend (or $TB$) phase~\cite{Dozov01}. 
Physically, the system undergoes a chiral spontaneous symmetry breaking. 
Although first predicted theoretically in~\cite{Dozov01}, 
the $TB$ phase was only discovered experimentally~\cite{Chen13,Borshch13,Wang15} and numerically~\cite{Memmer02} recently.
In fact in three-dimension, the $TB$ phase is competing with the $SB_\parallel$ phase, depending on the ratio of the twist to the splay elastic constant.

In this paper, we will consider \emph{polar} pizza-slice-shaped molecules which can spontaneously splay (see Fig.~\ref{fig:banana-pizza}(b)) and
\emph{polar} banana molecules which can spontaneously bend and align (see Fig.~\ref{fig:polar-banana}(b)).
In particular, we discover two new additional splay-bend phases, which we call $SB_\perp$ and $SB_\infty$.
(Other molecular shapes such as flag-shaped have also been considered in~\cite{Hinshaw88} and 
they can also give rise to modulated phases such as cubic and hexagonal phases~\cite{Blankschtein85}, which are different from ours.)

To model spontaneous bend, 
one can allow the bend elastic constant $K_3$ in the Frank elastic energy $H[\hat{\boldsymbol{n}}(\boldsymbol{r})]$ to be negative.
To prevent $H[\hat{\boldsymbol{n}}(\boldsymbol{r})]$ from going to $-\infty$, one also has to add higher order terms in gradient to stabilise it~\cite{Dozov01}.
(It turns out that only one such term is sufficient to bound the energy density from below, see Section I~\cite{supp}.)
Another approach is to introduce two fields to indicate the director field and the bend direction~\cite{Virga14}
or fast and slowly-varying fields~\cite{Kats14}.
Here, we extend the former approach by allowing the splay elastic constant $K_1$ to be negative to model spontaneous splay.

For example in pizza-slice-shaped molecules, 
one can imagine these molecules to be ferroelectric such that at low enough temperature the molecules tend to align in the same polarization
and this induces a local splay deformation.
Consequently, liquid crystals formed by pizza-slice-shaped molecules are also \emph{polar}~\cite{Cates18}.
In polar liquid crystals, we also have polarization field to indicate the local average polarization of the molecules in addition to director field which only indicates the orientation of the molecules.
However in our case, the configuration of the director field uniquely defines the polarization field.
For instance in polar pizza-slices, the polarization field always points in the direction of splay: $\propto(\nabla\cdot\hat{\mathbf{n}})\hat{\mathbf{n}}$.

It should be noted that our model is different from flexoelectric effect in liquid crystals~\cite{Meyer69,Lonberg85}.
In the case of flexoelectricity, the splay is induced by an external electric field $\mathbf{E}$ (see Fig.~\ref{fig:banana-pizza}(c))
and thus the molecules only tend to splay in the direction of $\mathbf{E}$.
Conversely, in our model, the splay is induced \emph{via} spontaneous symmetry breaking.
For instance, in Fig.~\ref{fig:banana-pizza}(b), the molecules can either splay to the left or to the right equally likely.
Incidentally, under confinement, strong anchoring at the walls can also induce spontaneous splay inside the liquid crystals~\cite{Barbero03}.
However this transition is more akin to Freedericks transition, \emph{i.e.}
\emph{global} spontaneous symmetry breaking as opposed to \emph{local}.

\begin{figure}[t]
\begin{centering}
\includegraphics[width=1\columnwidth]{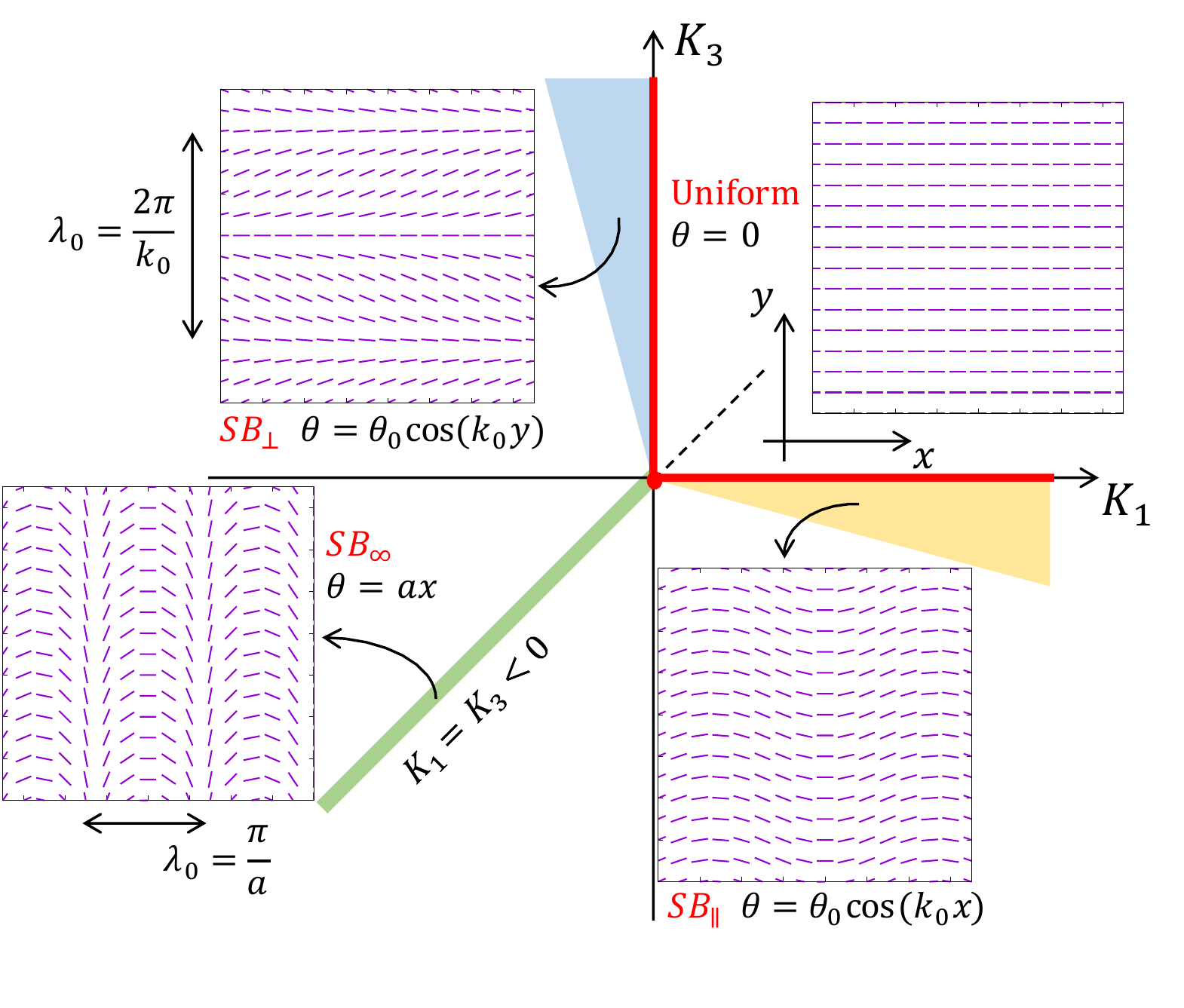}
\par\end{centering}
\caption{
Mean field phase diagram in the $K_{1}$-$K_{3}$ parameter space. 
We identify four distinct phases:
1) uniform nematic ($K_{1}$ and $K_{3}>0$),
2) $SB_{\perp}$ (blue triangular region in second quadrant), 
3) $SB_{\parallel}$ (yellow triangular region in the third quadrant), and
4) $SB_{\infty}$ phase (on the line $K_{1}=K_{3}<0$).
The four phases are separated by second order transition lines (red lines). 
The insets show the director field configuration $\hat{\mathbf{n}}(x,y)$ for each phase. \label{fig:phase-diagram}}
\end{figure}

Finally, we will consider \emph{polar} banana-shaped molecules.
In each molecule, we add a tiny electric polarization in the direction perpendicular to the longest molecular axis 
(red arrows in Fig.~\ref{fig:polar-banana}(b)).
Obviously polar banana molecules are also flexoelectric~\cite{Meyer69} but we do not consider the effects of external field or boundaries here.
Instead, we will only consider genuine phase transition from the uniformly nematic phase (where $\hat{\mathbf{n}}(\mathbf{r})$ is constant) into a completely new phase, 
$SB_\infty$, as shown in Fig.~\ref{fig:polar-banana}(b).
The director field $\hat{\mathbf{n}}(\mathbf{r})$ corresponding to the $SB_\infty$ phase is shown in bottom left of Fig.~\ref{fig:phase-diagram}.
The polarization field, on the other hand, points in the direction of bend: $\propto\hat{\mathbf{n}}\times\nabla\times\hat{\mathbf{n}}$.
Thus the $SB_\infty$ phase acquires a macroscopic polarization in some randomly chosen direction, which is upwards in the figure.
However, the phase transition from the uniform nematic (zero net polarization) to the $SB_\infty$ phase (macroscopic polarization) is not simply described by para-ferromagnetic transition due to an additional constraint in the system, which we shall see later.

In this paper, we shall consider an effective field theory for the director field $\hat{\mathbf{n}}(\mathbf{r})$
(since the polarization field can be determined from $\hat{\mathbf{n}}(\mathbf{r})$, if needed).
We shall also restrict to a two-dimensional director field and spatial dimension $d=2$ or $3$.
In the case of $d=3$, physically, the molecules form a columnar stack in the $z$-direction (see Fig.~\ref{fig:banana-pizza}(d))~\cite{Miyajima12}.


We will now derive analytically the mean field phase diagram as a function of splay ($K_1$) and bend ($K_3$) elastic constants, as shown in Fig.~\ref{fig:phase-diagram}.
(Note that $K_1$ and $K_3$ should be interpreted as effective elastic constants which account for steric repulsions and ferroelectric interactions.)
We start from the Frank elastic energy $H[\hat{\mathbf{n}}(\mathbf{r})]=\int d^{d}r\,f$,
where the energy density $f$ is given by the gradient expansion in $\hat{\mathbf{n}}(\mathbf{r})$~\cite{deGennes,Dozov01}:
\begin{equation}
f = \frac{K_{1}}{2}(\nabla\cdot\hat{\mathbf{n}})^{2} + \frac{K_{3}}{2}\left|\hat{\mathbf{n}}\times\nabla\times\hat{\mathbf{n}}\right|^{2}+\frac{C}{2}(\nabla^{2}\hat{\mathbf{n}})^{2}, \label{eq:f-n}
\end{equation}
and $|\hat{\mathbf{n}}(\mathbf{r})|=1$.
We require $C>0$ for stability (see Section I~\cite{supp}), but $K_{1}$ and $K_{3}$ can be negative.
In the case of $K_{1}<0$ and/or $K_{3}<0$, the director field locally acquires a spontaneous splay and/or bend (like pizza-slices or bananas).
(We assume there is no twist for simplicity.)

Since $\hat{\mathbf{n}}(\mathbf{r})$ is a two-dimensional vector, we can write: $\hat{\mathbf{n}}(\mathbf{r})=(\cos\theta(\mathbf{r)},\sin\theta(\mathbf{r)})^{T}$,
where $\theta(\mathbf{r})$ is the angle between the director field and the $x$-axis. 
The energy density (\ref{eq:f-n}) then becomes:
\begin{align}
f & =\frac{K_{1}+K_{3}}{4}\left|\nabla\theta\right|^{2}+\frac{C}{2}(\nabla^{2}\theta)^{2}+\frac{C}{2}\left|\nabla\theta\right|^{4}\nonumber \\
 & -\frac{K_{1}-K_{3}}{4}\left[\left(\frac{\partial\theta}{\partial x}\right)^{2}-\left(\frac{\partial\theta}{\partial y}\right)^{2}\right]\cos(2\theta)\nonumber \\
 & -\frac{K_{1}-K_{3}}{2}\left(\frac{\partial\theta}{\partial x}\right)\left(\frac{\partial\theta}{\partial y}\right)\sin(2\theta).\label{eq:f-theta}
\end{align}
The last two terms in the above equation can also be written as:
\begin{equation}
-\frac{K_{1}-K_{3}}{2}(\nabla\theta)^{T}\cdot\underline{\underline{\mathbf{Q}}}\cdot\nabla\theta, \label{eq:K1-K3-term}
\end{equation}
where 
\begin{equation}
\underline{\underline{\mathbf{Q}}} = \frac{1}{2} 
\left(\begin{array}{cc}
\cos(2\theta) & \sin(2\theta)\\
\sin(2\theta) & -\cos(2\theta)
\end{array}\right).
\end{equation}
Under some two-dimensional rotation $\underline{\underline{\mathbf{R}}}$, $\underline{\underline{\mathbf{Q}}}$ transforms as 
$\underline{\underline{\mathbf{Q}}}\rightarrow\underline{\underline{\mathbf{R}}}\cdot\underline{\underline{\mathbf{Q}}}\cdot\underline{\underline{\mathbf{R}}}^{T}$
and $\nabla\theta$ transforms as 
$\nabla\theta\rightarrow\underline{\underline{\mathbf{R}}}\cdot\nabla\theta$.
Therefore (\ref{eq:K1-K3-term}), and consequently (\ref{eq:f-theta}), is invariant under two-dimensional rotation as required from $H[\hat{\mathbf{n}}(\mathbf{r})]$ (\ref{eq:f-n}).

From the mean field phase diagram in Fig.~\ref{fig:phase-diagram},
we can identify four distinct phases: uniform nematic, $SB_{\perp}$, $SB_{\parallel}$, and $SB_{\infty}$,
separated by critical lines or second order phase transitions (red lines in the figure).
The case of $SB_{\parallel}$ has been reported before but not $SB_{\perp}$ or $SB_{\infty}$.
In the first quadrant (\emph{i.e.} $K_{1}>0$ and $K_{3}>0$), 
we have the uniform phase where $\theta(\mathbf{r})$ is constant everywhere in space and 
thus the director field $\hat{\mathbf{n}}(\mathbf{r})$ is pointing along some spontaneously-broken direction, which is the $x$-direction in the figure.
This corresponds to the usual nematic phase formed by rod-shaped molecules.

For $K_{3}<0$ and $K_{1}\gg\left|K_{3}\right|$, which is approximately the yellow triangular region in Fig.~\ref{fig:phase-diagram}, we have the $SB_\parallel$ phase. 
This corresponds to \emph{apolar} banana-shaped molecules, shown in Fig.~\ref{fig:banana-pizza}(a) and \ref{fig:polar-banana}(a).
In this phase, the director field $\hat{\mathbf{n}}(\mathbf{r})$ oscillates in the direction parallel to the global director $\bar{\mathbf{n}}=\frac{1}{V}\int\hat{\mathbf{n}}(\mathbf{r})dV$
(see bottom right inset in Fig.~\ref{fig:phase-diagram}). 
In other words, from uniform to $SB_\parallel$ phase, translational symmetry along the global director $\bar{\mathbf{n}}$ is broken.
In this paper, we choose the spontaneously broken direction to be $\bar{\mathbf{n}}=\hat{x}$, and thus in the $SB_\parallel$ phase,  
$\hat{\mathbf{n}}(\mathbf{r})$ oscillates along the $x$-axis. 
Mathematically, the mean field solution to the $SB_{\parallel}$ phase can be approximated as $\theta(\mathbf{r})=\theta_{0}\cos(k_{0}x)$,
where $\theta_{0}$ and $k_{0}$ depend on $K_{1}$ and $K_{3}$. 

For $K_{1}<0$ and $K_{3}\gg|K_{1}|$, or the blue triangular region in Fig.~\ref{fig:phase-diagram}, we have the $SB_\perp$ phase. 
This phase is formed by pizza-slice-shaped molecules, shown in Fig.~\ref{fig:banana-pizza}(b).
In this phase, $\hat{\mathbf{n}}(\mathbf{r})$ oscillates in the direction perpendicular to $\bar{\mathbf{n}}$, which is along $\hat{y}$ (see top left inset in Fig.~\ref{fig:phase-diagram}). Mathematically, the mean field solution to $SB_{\parallel}$ can be approximately as $\theta(\mathbf{r})=\theta_{0}\cos(k_{0}y)$.

Finally along the line $K_{1}=K_{3}<0$, or the green line in Fig.~\ref{fig:phase-diagram}, we have the $SB_{\infty}$ phase. 
In this phase, $\hat{\mathbf{n}}(\mathbf{r})$ tumbles along some spontaneously broken direction, which is $\hat{x}$ in the figure. 
Mathematically the mean field solution to this phase is given \emph{exactly} by $\theta(\mathbf{r})=ax+by$ for some constants $a$ and $b$ which depend on $K_{1}$ and $K_{3}$
($b=0$ in Fig.~\ref{fig:phase-diagram}).
This corresponds roughly to \emph{polar} banana molecules shown in Fig.~\ref{fig:polar-banana}(b).

\begin{table*}[t]
\begin{centering}
\begin{tabular}{c|c|c}
 & Para-ferromagnetic transition & Uniform nematic-$SB_{\infty}$ transition 
\tabularnewline
\hline 
Order parameter: & 
$\mathbf{m}(\mathbf{r}):\mathbb{R}^{d}\rightarrow\mathbb{R}^{n}$ & 
$\mathbf{m}(\mathbf{r})=\nabla\theta(\mathbf{r}):\mathbb{R}^{d}\rightarrow\mathbb{R}^{d}$
\tabularnewline
Hamiltonian: & 
$H[\mathbf{m}]=\int d^{d}r\left\{ \frac{K}{2}\left|\mathbf{m}\right|^{2}+\frac{B}{4}\left|\mathbf{m}\right|^{4}+\frac{C}{2}\left|\nabla\mathbf{m}\right|^{2}\right\} $ & $
H[\mathbf{m}]=\int d^{d}r\left\{ \frac{K}{2}\left|\mathbf{m}\right|^{2}+\frac{B}{4}\left|\mathbf{m}\right|^{4}+\frac{C}{2}\left|\nabla\cdot\mathbf{m}\right|^{2}\right\} $
\tabularnewline
 &  & and $\nabla\times\mathbf{m}=0$
\tabularnewline
Critical exponent for & &
\tabularnewline
$\xi\sim\left|K-K_{c}\right|^{-\nu}$: & 
$\nu=\frac{1}{2}+\frac{n+2}{8(n+8)}\epsilon+\mathcal{O}(\epsilon^{2})$, where $\epsilon=4-d$ & 
$\nu=\frac{1}{2}+\frac{3}{20}\epsilon+\mathcal{O}(\epsilon^{2})$, where $\epsilon=4-d$
\end{tabular}
\par\end{centering}
\caption{
Comparison between our uniform nematic-to-$SB_{\infty}$ transition (right column) and para-to-ferromagnetic transition (left column).
We show that uniform nematic to $SB_{\infty}$ phase transition belongs to a different universality class due to the constraint $\nabla\times\mathbf{m}=0$~\cite{curl} 
for our order parameter $\mathbf{m}(\mathbf{r})=\nabla\theta(\mathbf{r})$ (here $B=2C$).  \label{table:RG}}
\end{table*}

First we shall look at the mean field transition from the uniform phase, where $\theta(\mathbf{r})=0$, to the $SB_{\perp}$ phase, where $\theta(\mathbf{r})=\theta_{0}\cos(k_{0}y)$. 
In other words, we fix $K_{3}$ to be a positive constant and we decrease $K_{1}$ slowly from a positive value to a negative value,
crossing the critical line $K_1=0$ (red line on the $K_{3}$-axis in Fig.~\ref{fig:phase-diagram}).
To characterize this transition, we substitute the solution $\theta(\mathbf{r})=\theta_{0}\cos(k_{0}y)$ to the Hamiltonian density (\ref{eq:f-theta}). 
We then average the Hamiltonian density over one wavelength: $\bar{f}=\frac{2\pi}{k_{0}}\int_{0}^{2\pi/k_{0}}f\,dy$ and the result is:
\begin{equation}
\bar{f}=\frac{1}{4}K_{1}k_{0}^{2}\theta_{0}^{2}+\frac{1}{16}(K_{3}-K_{1})k_{0}^{2}\theta_{0}^{4}+\frac{1}{4}Ck_{0}^{4}\theta_{0}^{2}+\mathcal{O}(k_{0}^{4}\theta_{0}^{4}).\label{eq:f-SB-perp}
\end{equation}
We then minimize the average Hamiltonian over $\theta_0$ and $k_0$: $\partial\bar{f}/\partial\theta_{0}=\partial\bar{f}/\partial k_{0}=0$,
to obtain the solutions for $\theta_{0}$ and $\lambda_{0}=2\pi/k_{0}$:
\begin{align}
\theta_{0} & = 
\begin{cases} 
	0 & ,\,\, K_{1}>0 \\
	\sqrt{\frac{4}{3}\frac{-K_{1}}{K_{3}-K_{1}}} & ,\,\, K_{1}<0
\end{cases} \label{eq:theta0} \\
\lambda_{0} = \frac{2\pi}{k_0} & =
\begin{cases}
	0 & ,\,\, K_{1}>0 \\
	2\pi\sqrt{\frac{3C}{-K_{1}}} & ,\,\, K_{1}<0.
\end{cases} \label{eq:k0}
\end{align}
Here $\lambda_{0}$ is the wavelength of the splay modulation (see top left inset in Fig.~\ref{fig:phase-diagram}). 
As we approach the critical line from below, the order parameter $\theta_0$ vanishes as $\theta_0\sim|K_1|^\beta$ with mean field exponent $\beta=1/2$,
whereas $\lambda_{0}$ becomes longer and longer.
Note that since we have neglected a higher order term $\propto k_{0}^{4}\theta_{0}^{4}$ in (\ref{eq:f-SB-perp}),
we require $k_{0}$ and $\theta_{0}$ to be small. 
From (\ref{eq:theta0}-\ref{eq:k0}), $k_{0}$ and $\theta_{0}$ are small as long as $\left|K_{1}\right|$ is small and $K_{3}\gg\left|K_{1}\right|$. 
This gives the blue triangular region in the top of Fig.~\ref{fig:phase-diagram}. 
Far from this region, the $SB_\perp$ phase is no longer accurately represented by $\theta(\mathbf{r})=\theta_{0}\cos(k_{0}y)$
and one may expect higher order harmonic terms in $\theta(\mathbf{r})$.
Similarly, the region between $SB_\perp$ and $SB_\infty$ and that between $SB_\parallel$ and $SB_\infty$ in the mean field phase diagram Fig.~\ref{fig:phase-diagram}
are not known analytically and one has to do extensive numerical simulations.


The mean field transition from uniform to $SB_\parallel$ can be calculated in similar fashion as above.
In fact, the equilibrium configurations $\hat{\mathbf{n}}(\mathbf{r})$ for $SB_{\perp}$ and $SB_{\parallel}$ are symmetric under transformation:
$\hat{n}_\alpha\rightarrow\epsilon_{\alpha\beta}\hat{n}_\beta$ and swapping $K_1\leftrightarrow K_3$ (Hodge duality).


Finally we consider the transition from the uniform to the $SB_{\infty}$ phase, as shown in Fig.~\ref{fig:polar-banana}(b). 
The $SB_{\infty}$ phase is located along the line $K_{1}=K_{3}<0$ in the mean field phase diagram, see green line in Fig.~\ref{fig:phase-diagram}.
The easiest way to characterize this transition is to assume $K_{1}=K_{3}=K$ (single elastic constant approximation).
The Hamiltonian~(\ref{eq:f-theta}) then becomes:
\begin{equation}
H[\theta(\mathbf{r})]=\int d^{d}r\left\{ \frac{K}{2}\left|\nabla\theta\right|^{2}+\frac{C}{2}(\nabla^{2}\theta)^{2}+\frac{C}{2}\left|\nabla\theta\right|^{4}\right\} .\label{eq:H-theta}
\end{equation}
We then define a vector field $\mathbf{m}(\mathbf{r})=\nabla\theta(\mathbf{r})$ and substituting this to (\ref{eq:H-theta}), 
we obtain the Hamiltonian in terms of $\mathbf{m}$:
\begin{equation}
H[\mathbf{m}]=\int d^{d}r\left\{ \frac{K}{2}\left|\mathbf{m}\right|^{2}+\frac{C}{2}\left|\mathbf{m}\right|^{4}+\frac{C}{2}(\nabla\cdot\mathbf{m})^{2}\right\} ,\label{eq:H-m}
\end{equation}
which looks like a ferromagnet except for the constraint $\nabla\times\mathbf{m}=0$~\cite{curl} (see Table~\ref{table:RG}).

Note that in para-ferromagnetic transition (left column, Table~\ref{table:RG}),
the dimension of the order parameter is $n$ whereas the spatial dimension is $d$.
Using renormalization group, the critical exponent $\nu$ can be given in terms of $\epsilon$-expansion from spatial dimension $d=4$.
Many physical systems fall into this broad universality class.
For instance, Ising model and liquid/gas critical point correspond to $n=1$ (up-down symmetry)
whereas XY-model corresponds to $n=2$~\cite{Kardar}.
On the other hand for our constrained Hamiltonian, the dimension of our order parameter $\mathbf{m}=\nabla\theta$ is equal to $d$ (right column, Table~\ref{table:RG}),
Moreover, we also get a different critical exponent $\nu$, indicating a different universality class from that of the unconstrained one.
(Note that we started from a director field $\hat{\mathbf{n}}$ is a two-dimensional vector,
and we mapped it to $\mathbf{m}$ which is $d$-dimensional.)

At mean field level, the solution for $\mathbf{m}$ to the Hamiltonian (\ref{eq:H-m}) is one which minimizes $H[\mathbf{m}]$:
$\delta H/\delta\mathbf{m}=0$, from which we obtain, in $d=2$:
\begin{equation}
\mathbf{m}(\mathbf{r})=\begin{cases}
\left(0,0\right)^{T} & ,\,\,K>0\\
\left(a,b\right)^{T} & ,\,\,K<0
\end{cases}
\end{equation}
where $a$ and $b$ satisfy $a^{2}+b^{2}=-K/2C$. Inverting
$\mathbf{m}$, we obtain $\theta$:
\begin{equation}
\theta(\mathbf{r}) = 
\begin{cases}
  \text{constant} & ,\,\, K>0 \\
  ax+by              & ,\,\, K<0
\end{cases}
\end{equation}
as expected. 
We can also calculate the fluctuations from the mean field: $\delta\mathbf{m}(\mathbf{r})=\mathbf{m}(\mathbf{r})-\mathbf{m}_0$,
and the correlation function $\left<\delta\mathbf{m}(\mathbf{r})\cdot\delta\mathbf{m}(\mathbf{r}')\right>$.
From the correlation function, we can extract the correlation length, which is given by $\xi=\sqrt{-C/K}$ for the $SB_\infty$ phase (see Section II of~\cite{supp}).
Thus the correlation length diverges at critical point $K=0$ with critical exponent $\nu=1/2$.
At mean field level, we cannot distinguish the critical exponent of our constrained Hamiltonian from the unconstrained ferromagnetic transition.
Furthermore, mean field calculation also predicts that the critical point is at $K=0$.

Renormalization group procedure allows us to get higher order correction to the mean field exponent $\nu=1/2$ (detailed in Section III of~\cite{supp}).
We show that at linear order in $\epsilon=4-d$, the critical exponent for the uniform-to-$SB_\infty$ transition is indeed different from that of unconstrained ferromagnetic transition
(see Table~\ref{table:RG}).
It is interesting to investigate if there are other physical systems which belong to the same universality class as ours.
Note that in $d=2$, all these phases (including the uniform nematic) become quasi long-range order.
In $d=2$, the transition from the isotropic to quasi-nematic phase is of Kosterlitz-Thouless type~\cite{Frenkel85,Kosterlitz74},
however, it is not clear if the same is true for the quasi-nematic to any of the quasi-$SB$ phases.

In conclusion, using field-theoretic methods, we showed that pizza-slice-shaped molecules and polar bananas can give rise to new exotic liquid-crystalline phases.
It might be interesting to generalize the above calculation to full three-dimension not just confined to two-dimensional layers, 
and hopefully, the existence of these phases can also be confirmed experimentally in the future.
It might also be interesting to compare our results to particle-based simulations such as Monte Carlo or Molecular Dynamics.

\begin{acknowledgments}
We thank M. E. Cates, J. Ball, C. Nardini, and F. Caballero for illuminating discussions. 
XM acknowledges support from Undergraduate Research Opportunity Bursary.
Work funded in part by the European Research Council under the Horizon 2020 Programme, ERC grant agreement number 740269.
\end{acknowledgments}

\newpage
\onecolumngrid

\section*{Supplementary Material}

\section{Proof That the Hamiltonian is Bounded from Below}

We shall show that the energy density
\begin{equation}
f(\mathbf{n}) = \frac{K_1}{2}(\nabla\cdot\mathbf{n})^2
  		      + \frac{K_2}{2}\left(\mathbf{n}\cdot(\nabla\times\mathbf{n})\right)^2
		      + \frac{K_3}{2}|\mathbf{n}\times(\nabla\times\mathbf{n})|^2
		      + \frac{C}{2}|\nabla^2\mathbf{n}|^2, \label{energy-density}
\end{equation}
subject to $|\mathbf{n}|^2=1$, is bounded below. We shall only consider the case where $K_1,K_2,K_3<0$ (and $C>0$ for stability), since otherwise we could simply drop the non-negative term.
The idea is to ``decouple'' $f$ into a sum of contributions from $n_\alpha$, where $\alpha=1,2,3$ represents Cartesian coordinates. We will need: \\

\textbf{Proposition} (Cauchy-Schwarz inequality). \textit{Let $\mathbf{v}$, $\mathbf{w}\in\mathbb{R}^n$. Then 
\begin{equation}
(\mathbf{v}\cdot\mathbf{w})^2\le|\mathbf{v}|^2|\mathbf{w}|^2,
\end{equation}
with equality iff one of $\mathbf{v}$, $\mathbf{w}$ is a multiple of the other.} \\

We bound the first term as follows:
\begin{align}
(\nabla\cdot\mathbf{n})^2 &= \left(\partial_1n_1+\partial_2n_2+\partial_3n_3\right)^2 \nonumber\\
					     &\le 3\left((\partial_1n_1)^2+(\partial_2n_2)^2+(\partial_3n_3)^2\right) \nonumber\\
					     &\le 3\left(|\nabla n_1|^2+|\nabla n_2|^2+|\nabla n_3|^2\right),
\end{align}
where when going to the second line we used Cauchy-Schwarz with $\mathbf{v}=(\partial_1n_1,\partial_2n_2,\partial_3n_3)$ and $\mathbf{w}=(1,1,1)$. Next,
\begin{align}
\left(\mathbf{n}\cdot(\nabla\times\mathbf{n})\right)^2 &\le |\nabla\times\mathbf{n}|^2 \nonumber\\
											    &= (\partial_2n_3-\partial_3n_2)^2+(\partial_3n_1-\partial_1n_3)^2+(\partial_1n_2-\partial_2n_1)^2 \nonumber\\
											    &\le 2\left((\partial_2n_3)^2+(\partial_3n_2)^2+(\partial_3n_1)^2+(\partial_1n_3)^2+(\partial_1n_2)^2+(\partial_2n_1)^2\right) \nonumber\\
											    &\le 2\left(|\nabla n_1|^2+|\nabla n_2|^2+|\nabla n_3|^2\right),
\end{align}
where again in the third line we used Cauchy-Schwarz. 
Note that we also have $|\mathbf{n}\times(\nabla\times\mathbf{n})|^2\le |\nabla\times\mathbf{n}|^2$, so exactly the same bound holds for the third term in (\ref{energy-density}).

Putting all of these together, we have
\begin{equation}
f \ge \sum_{\alpha=1}^3\left[ \left(\frac{3}{2}K_1+K_2+K_3\right)|\nabla n_\alpha|^2 + \frac{1}{2}C(\nabla^2 n_\alpha)^2 \right]. \label{bound_1}
\end{equation}
Observe the $\ge$ sign due to the assumption that $K_i<0$ for all $i$.

Now the trick is to add a constant term $A|\mathbf{n}|^4=A$ to $f$ for some constant $A>0$. This does not affect whether or not the free energy is bounded below. Using $|\mathbf{n}|^4\ge n_1^4+n_2^4+n_3^4$ and (\ref{bound_1}), we have
\begin{align}
\tilde{f} &= f+A \nonumber\\
	     &= f+A|\mathbf{n}|^4 \nonumber\\
	     &\ge \sum_{\alpha=1}^3\left[A\,n_\alpha^4 + \left( \frac{3}{2}K_1 + K_2 + K_3 \right)|\nabla n_\alpha|^2 + \frac{1}{2}C(\nabla^2 n_\alpha)^2\right].
\end{align}
Thus we have reduced the problem to showing that the new free energy 
\begin{equation}
g(\phi)=A\phi^4+\left(\frac{3}{2}K_1+K_2+K_3\right)|\nabla\phi|^2+\frac{1}{2}C(\nabla^2\phi)^2
\end{equation}
is bounded below. After rescaling $\mathbf{r}$ and $\phi$, this becomes
\begin{equation}
g(\phi)=\phi^4+2B|\nabla\phi|^2+(\nabla^2\phi)^2 \label{free_energy_2}
\end{equation}
for some constant $B$. 
Note that we started with the constraint $|\mathbf{n}|^2=1$, but at this stage we can drop this constraint and take $\phi\in\mathbb{R}$. 
Assuming that the boundary term is zero, we can integrate by parts and then complete the square:
\begin{align}
g &= \phi^4-2B\phi\nabla^2\phi+(\nabla^2\phi)^2 \nonumber\\
   &= \left(\nabla^2\phi-B\phi\right)^2-B^2\phi^2+\phi^4 \nonumber\\
   &= \left(\nabla^2\phi-B\phi\right)^2+\left(\phi^2-\frac{B^2}{2}\right)^2-\frac{B^4}{4},
\end{align}
which is now clearly bounded below.

\section{Structure factor and correlation length of the $SB_\infty$ phase}
The mean field solution to the $SB_\infty$ phase is given by:
\begin{equation}
\mathbf{m}_0 = \sqrt{\frac{-K}{2C}}\hat{x},
\end{equation}
where $K<0$ and we have chosen the spontaneously broken direction to be the $x$-direction.
Now let us consider some small fluctuation $\delta\mathbf{m}(\mathbf{r})$ around the mean field solution $\mathbf{m}_0$.
Substituting $\mathbf{m}(\mathbf{r})=\mathbf{m}_0+\delta\mathbf{m}(\mathbf{r})$ to the Hamiltonian, Eq. (9) in the main text, we obtain:
\begin{equation}
H[\delta\mathbf{m}]=\int d^dr \left\{ \frac{C}{2}\left(\nabla\cdot\delta\mathbf{m}\right)^2 - K\delta m_x^2 \right\},
\end{equation}
subject to constraint $\nabla\times\delta\mathbf{m}=0$.
In Fourier space,
\begin{equation}
\delta\mathbf{m}(\mathbf{r})=\int\frac{d^dq}{(2\pi)^{d/2}} \delta\mathbf{m}(\mathbf{q}) e^{i\mathbf{q}\cdot\mathbf{r}},
\end{equation}
this Hamiltonian becomes:
\begin{equation}
H[\delta\mathbf{m}(\mathbf{q})] = \frac{1}{2} \int d^dq \left(Cq_\alpha q_\beta - 2K\delta_{x\alpha}\delta_{x\beta} \right) \delta m_\alpha(\mathbf{q})\delta m_\beta(-\mathbf{q}).
\end{equation}
In particular, in spatial dimension $d=2$, this becomes:
\begin{eqnarray}
H[\delta\mathbf{m}(\mathbf{q})] = \frac{1}{2} \int d^2q & & \left( Cq_x^2 - 2K\right)\delta m_x(\mathbf{q})\delta m_x(-\mathbf{q})
							      + Cq_x q_y \delta m_x(\mathbf{q})\delta m_y(-\mathbf{q})
							      + Cq_x q_y \delta m_y(\mathbf{q})\delta m_x(-\mathbf{q}) \nonumber\\
											    & &+ Cq_y^2  \delta m_y(\mathbf{q})\delta m_y(-\mathbf{q}).
\end{eqnarray}
Next we apply the constraint $\delta m_y(\mathbf{q})=\frac{q_y}{q_x}\delta m_x(\mathbf{q})$ to get:
\begin{equation}
H[\delta\mathbf{m}(\mathbf{q})] = \frac{1}{2} \int d^2q \left( Cq_x^2 + 2Cq_y^2 + C\frac{q_y^4}{q_x^2} -2K \right) \left|\delta m_x(\mathbf{q})\right|^2.
\end{equation}
This Hamiltonian is positive definite since $K<0$.
Therefore the structure factor is:
\begin{eqnarray}
S(\mathbf{q}) &=& \left<\left|\delta m_x(\mathbf{q})\right|^2\right> = \frac{1}{Cq_x^2 + 2Cq_y^2 + C\frac{q_y^4}{q_x^2} -2K} \\
  			 &=& \frac{1}{-K} \frac{1}{\xi^2q_x^2 + 2\xi^2q_y^2 + \xi^2\frac{q_y^4}{q_x^2} + 2} \\
			 &=& \frac{1}{-K} F(\xi\mathbf{q})
\end{eqnarray}
where $\xi=\sqrt{-C/K}$ is the correlation length and $F$ is some function independent of $C$ or $K$. 
Therefore the correlation length diverges as $K\rightarrow0^-$ as a power law:
$\xi\sim\left|K\right|^{-\nu}$,
with critical exponent $\nu=1/2$.

\section{Renormalization Group Along the Diagonal $K_{1}=K_{3}$ Line}

At mean field level, the critical exponent for the correlation length is $\nu=\frac{1}{2}$. 
Renormalization group allows us to get a more accurate estimate of the critical exponent. 
Usually, the critical exponent is given in expansion of $\epsilon$:
\begin{equation}
\nu=\frac{1}{2}+\alpha_1\epsilon+\alpha_2\epsilon^2+\ldots
\end{equation}
where $\epsilon=4-d$ is the distance from the upper critical dimension and $\alpha_i$'s are some constants. 
For $d\ge4$, mean field theory is exact and the critical exponent is exactly $\nu=\frac{1}{2}$. 
However for spatial dimension $d<4$, we have correction of order $\epsilon$ and higher.
In this section we will obtain the first order $\epsilon$ correction for the unifom to $SB_{\infty}$ transition along the diagonal line $K_{1}=K_{3}$ in the phase diagram.

The order parameter describing the uniform to $SB_{\infty}$ phase transition is a vector field $\mathbf{m}$ plus a constraint $\nabla\times\mathbf{m}=0$.
It is rather complicated to enforce this constraint to higher spatial dimension and thus we choose to work with a scalar field $\theta$ where $\mathbf{m}=\nabla\theta$.
First we write the Hamiltonian for $\theta$, Eq.~(8) in the main text, as follows:
\begin{equation}
H[\theta]=\int d^{d}r\Bigg\{\underbrace{\frac{K}{2}\left|\nabla\theta\right|^{2}+\frac{C}{2}(\nabla^{2}\theta)^{2}}_{\text{Gaussian }H_{0}}+\underbrace{L\left|\nabla\theta\right|^{4}}_{H_{I}}\Bigg\}
\end{equation}
where we have introduced a new constant $L$ for convenience. 
Later we can set $L=C/2$. 
The Hamiltonian above contains a Gaussian part, which we call $H_{0}[\theta]$, and a quartic part, which we call
$H_{I}[\theta]$. In Fourier space,
\begin{equation}
\theta(\mathbf{r})=\int\frac{d^{d}q}{(2\pi)^{d/2}}\,\theta_{\mathbf{q}}e^{i\mathbf{q}\cdot\mathbf{r}},
\end{equation}
the Hamiltonian can be written as:
\begin{align}
H_{0}[\theta_{\mathbf{q}}]= & \frac{1}{2}\int d^{d}q\left(Kq^{2}+Cq^{4}\right)\left|\theta_{\mathbf{q}}\right|^{2}\\
H_{I}[\theta_{\mathbf{q}}]= & L\int\frac{d^{d}q_{1}d^{d}q_{2}d^{d}q_{3}d^{d}q_{4}}{(2\pi)^{d}}\delta(\mathbf{q}_{1}+\mathbf{q}_{2}+\mathbf{q}_{3}+\mathbf{q}_{4})(\mathbf{q}_{1}\theta_{\mathbf{q}_{1}}\cdot\mathbf{q}_{2}\theta_{\mathbf{q}_{2}})(\mathbf{q}_{3}\theta_{\mathbf{q}_{3}}\cdot\mathbf{q}_{4}\theta_{\mathbf{q}_{4}})
\end{align}
The Gaussian part is exactly solvable, in particular, the correlation
function is given by (in the Gaussian limit $L\rightarrow0$):
\begin{equation}
\left\langle \theta_{\mathbf{q}}\theta_{\mathbf{q}'}\right\rangle _{0}=\frac{\delta(\mathbf{q}+\mathbf{q}')}{Kq^{2}+Cq^{4}}.\label{eq:correlation-theta}
\end{equation}

\begin{figure}
\begin{centering}
\includegraphics[width=0.6\columnwidth]{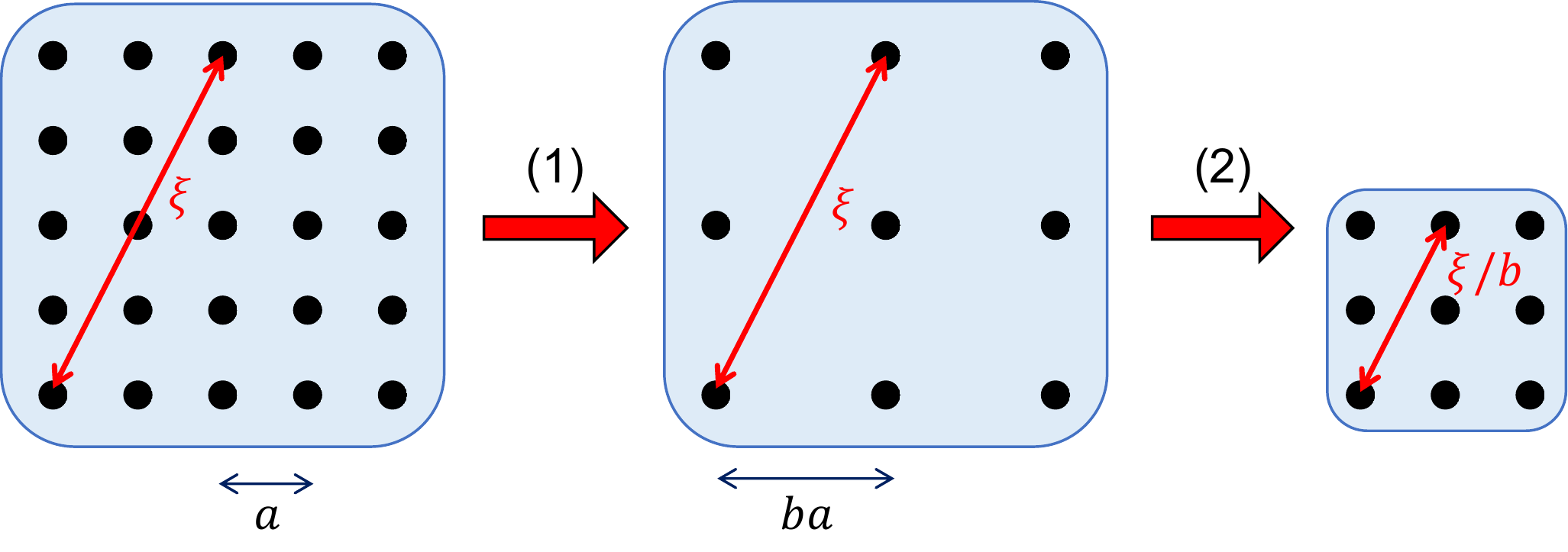}
\par\end{centering}
\caption{
Suppose the field is confined to a lattice with spacing $a$. 
The renormalization group procedure is as follows: 
(1) First we coarse-grain over some volume $(ba)^{d}$, where $b>1$. 
(2) Next we rescale the lattice positions $\mathbf{r}\rightarrow\mathbf{r}/b$ and the field itself $\phi\rightarrow\phi/z$. 
At the end of this procedure, the correlation length in the system is rescaled to $\xi\rightarrow\xi/b$.
\label{fig:RG}}
\end{figure}

The renormalization group (RG) consists of two steps. Suppose our
field $\theta(\mathbf{r})$ is confined to a lattice with lattice
spacing $a$. This defines an upper cut-off frequency: $\Lambda=2\pi/a$.
The first step of RG is the coarse-graining step: we average the field
over some box size $ba$, where $b>1$ (see Fig.~\ref{fig:RG}).
After coarse-graining, we end up with fewer lattice points. The second
step of RG is to rescale space $\mathbf{r}\rightarrow\mathbf{r}/b$
and the field itself $\phi\rightarrow\phi/z$. At the end of this
procedure, the correlation length $\xi$ is rescaled into $\xi/b$.
We can then repeat this procedure many times. This RG procedure turns
out to be useful when we are close to criticality, or $2^{\text{nd}}$ order
phase transition, where the correlation length $\xi$ is infinite.

First, in the coarse-graining step, we decompose the field $\theta_{\mathbf{q}}$
into low frequency and high frequency modes:
\begin{equation}
\theta_{\mathbf{q}}=\begin{cases}
\tilde{\theta}_{\mathbf{q}} & ;\,\,0<\left|\mathbf{q}\right|<\frac{\Lambda}{b}\\
\phi_{\mathbf{q}} & ;\,\,\frac{\Lambda}{b}\le\left|\mathbf{q}\right|<\Lambda
\end{cases}.
\end{equation}
The partition function is ($\beta=1$):
\begin{equation}
Z=\int\mathcal{D}\theta_{\mathbf{q}}\,e^{-H_{0}[\theta_{\mathbf{q}}]}e^{-H_{I}[\theta_{\mathbf{q}}]}.
\end{equation}
Substituting $\theta_{\mathbf{q}}=\tilde{\theta}_{\mathbf{q}}+\phi_{\mathbf{q}}$
and since we have: $H_{0}[\tilde{\theta}_{\mathbf{q}}+\phi_{\mathbf{q}}]=H_{0}[\tilde{\theta}_{\mathbf{q}}]+H_{0}[\phi_{\mathbf{q}}]$,
the partition function becomes:
\begin{align}
Z & =\int\mathcal{D}\tilde{\theta}_{\mathbf{q}}\,e^{-H_{0}[\tilde{\theta}_{\mathbf{q}}]}Z_{\phi}\int\mathcal{D}\phi_{\mathbf{q}}\,e^{-H_{I}}\frac{e^{-H_{0}[\phi_{\mathbf{q}}]}}{Z_{\phi}}\nonumber \\
 & =\int\mathcal{D}\tilde{\theta}_{\mathbf{q}}\,e^{-H_{0}[\tilde{\theta}_{\mathbf{q}}]}Z_{\phi}\left\langle e^{-H_{I}[\tilde{\theta}_{\mathbf{q}}+\phi_{\mathbf{q}}]}\right\rangle _{\phi}\nonumber \\
 & =Z_{\phi}\int\mathcal{D}\tilde{\theta}_{\mathbf{q}}\,e^{-H_{0}[\tilde{\theta}_{\mathbf{q}}]-\left\langle H_{I}\right\rangle _{\phi}+\frac{1}{2}\left(\left\langle H_{I}^{2}\right\rangle _{\phi}-\left\langle H_{I}\right\rangle _{\phi}^{2}\right)+\ldots}
\end{align}
where $\left\langle \cdot\right\rangle _{\phi}$ indicates averaging
over high frequency modes and $Z_{\phi}=\int\mathcal{D}\phi_{\mathbf{q}}\,e^{-H_{0}[\phi_{\mathbf{q}}]}$
is a constant. Defining the coarse-grained Hamiltonian to be $Z=\int\mathcal{D}\tilde{\theta}_{\mathbf{q}}\,e^{-\tilde{H}[\tilde{\theta}_{\mathbf{q}}]}$,
we get:
\begin{equation}
\tilde{H}[\tilde{\theta}_{\mathbf{q}}]=H_{0}[\tilde{\theta}_{\mathbf{q}}]+\underbrace{\left\langle H_{I}[\tilde{\theta}_{\mathbf{q}},\phi_{\mathbf{q}}]\right\rangle _{\phi}}_{{\cal O}(L)}-\frac{1}{2}\underbrace{\left(\left\langle H_{I}^{2}\right\rangle _{\phi}-\left\langle H_{I}\right\rangle _{\phi}^{2}\right)}_{{\cal O}(L^{2})}+\ldots.\label{eq:H-coarse-grain}
\end{equation}
We can ignore the constant term $\ln(Z_{\phi})$ which does not depend
on $\tilde{\theta}_{\mathbf{q}}$. The first term in (\ref{eq:H-coarse-grain})
is just the Gaussian part:
\begin{equation}
H_{0}[\tilde{\theta}_{\mathbf{q}}]=\frac{1}{2}\int_{0}^{\Lambda/b}d^{d}q\left(Kq^{2}+Cq^{4}\right)\left|\tilde{\theta}_{\mathbf{q}}\right|^{2}
\end{equation}
(note that the integration range is from $0$ to $\Lambda/b$). 

\begin{table*}[t]
\begin{centering}
\begin{tabular}{|c|c|c|}
\hline 
$L_{1}^{(1)}$ & $L_{2}^{(1)}$ & $L_{3}^{(1)}$\tabularnewline
\hline 
$1\times$\includegraphics[scale=0.4]{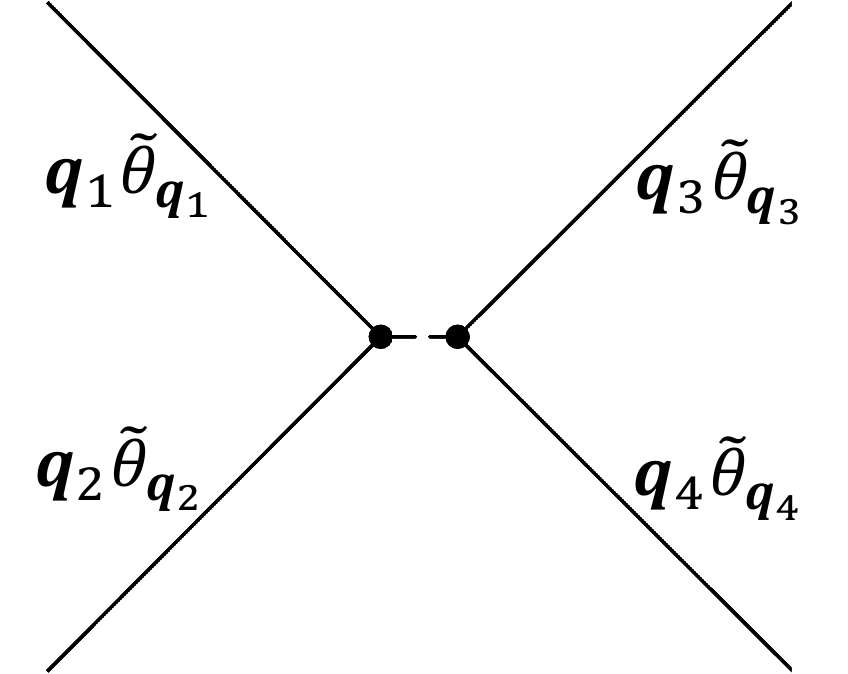} & $2\times$\includegraphics[scale=0.4]{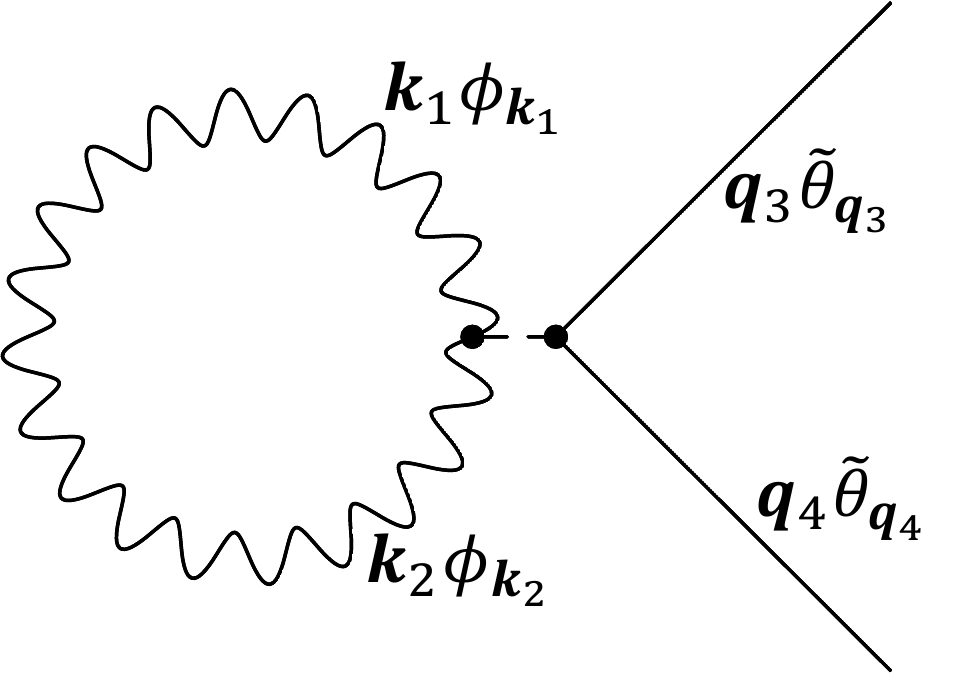} & $4\times$\includegraphics[scale=0.4]{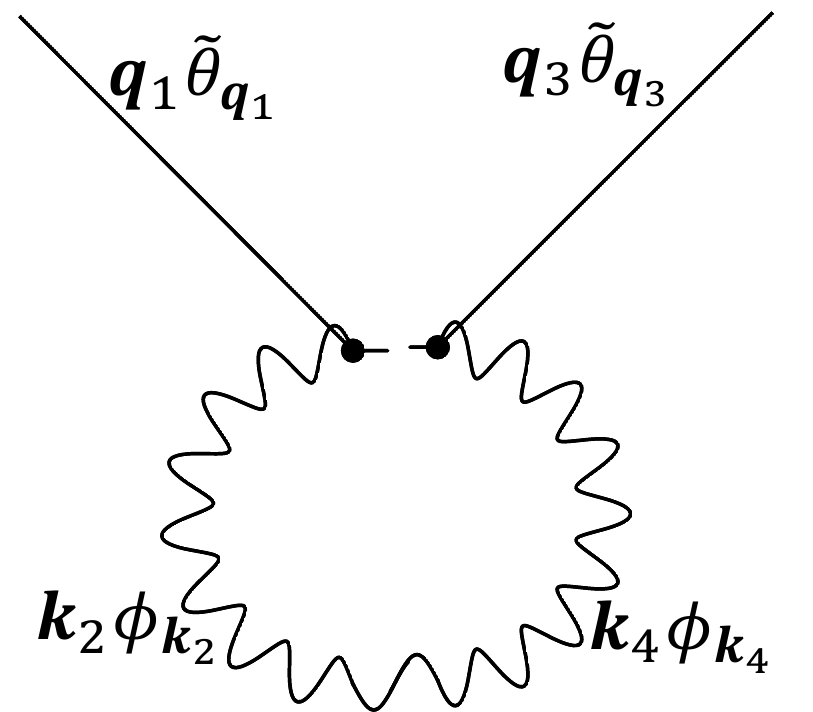}\tabularnewline
\hline 
$L_{4}^{(2)}$ & $L_{5}^{(2)}$ & $L_{6}^{(2)}$\tabularnewline
\hline 
$8\times$\includegraphics[scale=0.4]{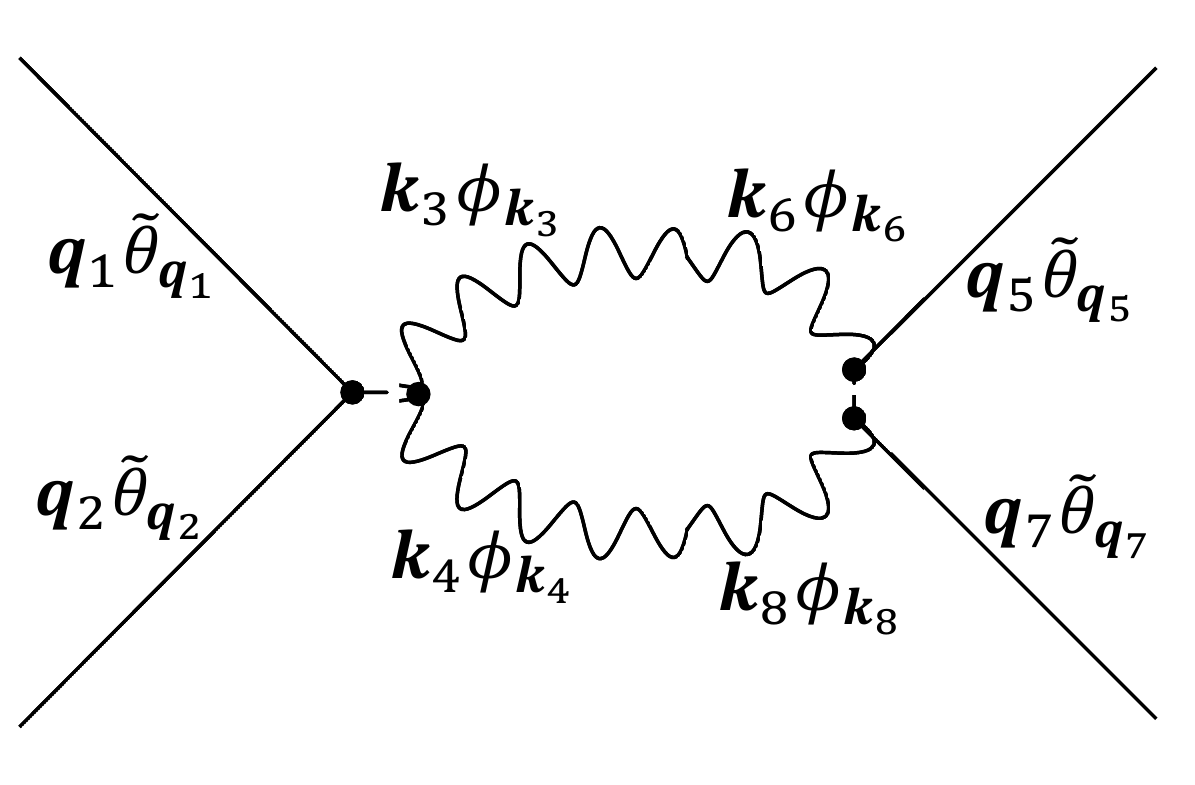} & $32\times$\includegraphics[scale=0.4]{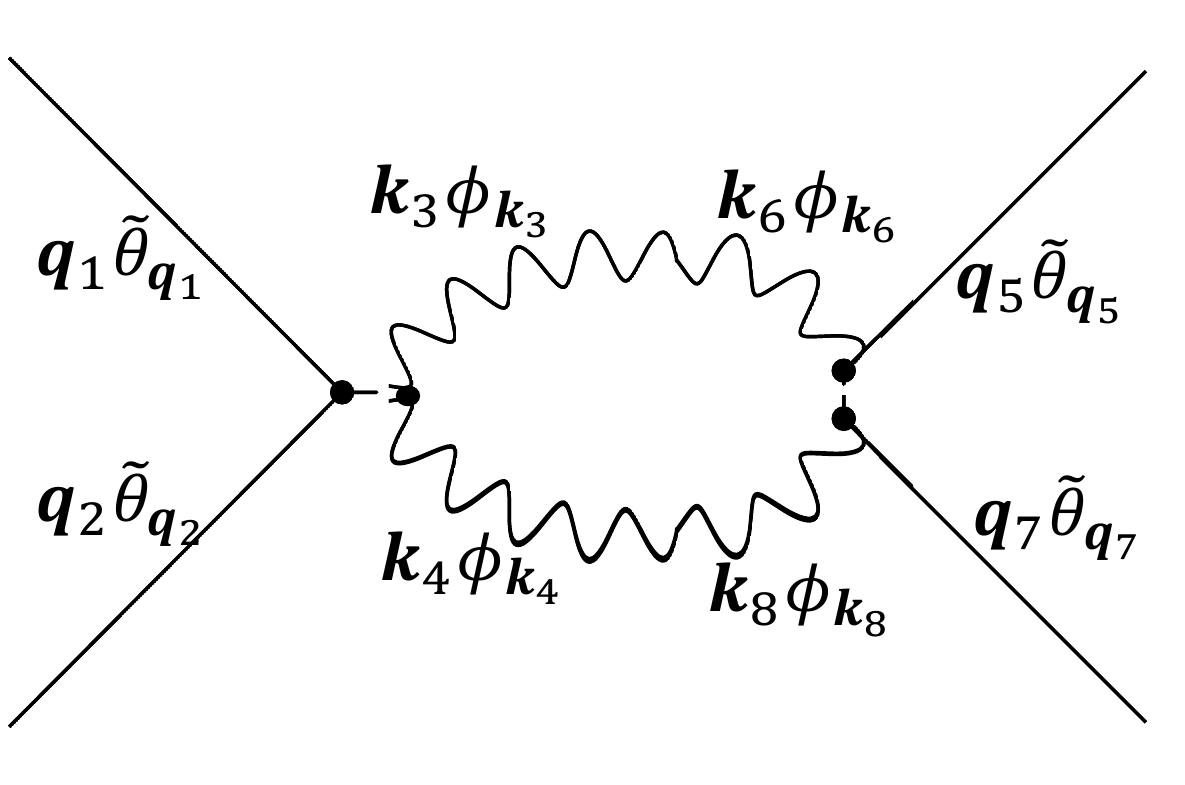} & $32\times$\includegraphics[scale=0.4]{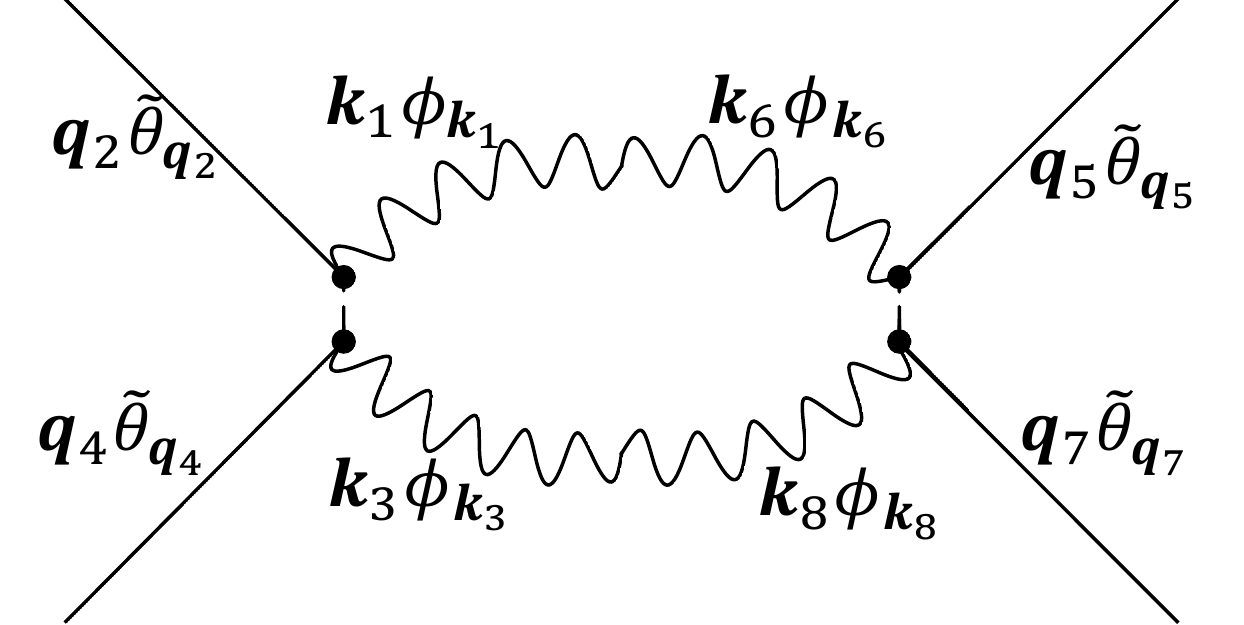}\tabularnewline
\hline 
\end{tabular}
\par\end{centering}
\caption{
Diagrammatic representation of the quartic term in the Hamiltonian: $\left\langle H_{I}\right\rangle -\frac{1}{2}\left(\left\langle H_{I}^{2}\right\rangle -\left\langle H_{I}\right\rangle ^{2}\right)$.
Solid lines represent low frequency modes, wavy lines represent high frequency modes, dots represent dot products, and
short dashed lines between two dots represent a momentum conservation,  \emph{e.g.} $\delta(\mathbf{k}_{1}+\mathbf{k}_{2}+\mathbf{q}_{3}+\mathbf{q}_{4})$ in $L_{2}^{(1)}$. 
(Note that we have followed diagrammatic convention of~\cite{Kardar}.) \label{tab:diagram}}
\end{table*}

\subsection{$\mathcal{O}(L)$ correction to $\tilde{H}[\tilde{\theta}_{\mathbf{q}}]$}

The second term in (\ref{eq:H-coarse-grain}) is the order ${\cal O}(L)$ correction:
\begin{align}
\left\langle H_{I}\right\rangle _{\phi} & =\frac{L}{2}\int\frac{d^{d}q_{1}\ldots d^{d}q_{4}}{(2\pi)^{d}}\delta(\mathbf{q}_{1}+\cdots\mathbf{q}_{4})(\mathbf{q}_{1}\cdot\mathbf{q}_{2})(\mathbf{q}_{3}\cdot\mathbf{q}_{4})\left\langle (\tilde{\theta}_{\mathbf{q}_{1}}+\phi_{\mathbf{q}_{1}})\cdot(\tilde{\theta}_{\mathbf{q}_{2}}+\phi_{\mathbf{q}_{2}})(\tilde{\theta}_{\mathbf{q}_{3}}+\phi_{\mathbf{q}_{3}})\cdot(\tilde{\theta}_{\mathbf{q}_{4}}+\phi_{\mathbf{q}_{4}})\right\rangle _{\phi}  \label{eq:order-L}\\
 & =L_{1}^{(1)}+L_{2}^{(1)}+L_{3}^{(1)}.
\end{align}
Expanding the integrand in (\ref{eq:order-L}), we may get terms such
as $\left\langle \phi\tilde{\theta}\tilde{\theta}\tilde{\theta}\right\rangle _{\phi}$,
which is zero since $\left\langle \phi\right\rangle _{\phi}=0$, or
$\left\langle \phi\phi\phi\phi\right\rangle _{\phi}$, which is constant.
The non-trivial terms are
\begin{align}
L_{1}^{(1)}= & L\int\frac{d^{d}q_{1}d^{d}q_{2}d^{d}q_{3}d^{d}q_{4}}{(2\pi)^{d}}\delta(\mathbf{q}_{1}+\mathbf{q}_{2}+\mathbf{q}_{3}+\mathbf{q}_{4})(\mathbf{q}_{1}\cdot\mathbf{q}_{2})(\mathbf{q}_{3}\cdot\mathbf{q}_{4})\tilde{\theta}_{\mathbf{q}_{1}}\tilde{\theta}_{\mathbf{q}_{2}}\tilde{\theta}_{\mathbf{q}_{3}}\tilde{\theta}_{\mathbf{q}_{4}}\label{eq:L1}\\
L_{2}^{(1)}=2\times & L\int\frac{d^{d}k_{1}d^{d}k_{2}d^{d}q_{3}d^{d}q_{4}}{(2\pi)^{d}}\delta(\mathbf{k}_{1}+\mathbf{k}_{2}+\mathbf{q}_{3}+\mathbf{q}_{4})(\mathbf{k}_{1}\cdot\mathbf{k}_{2})(\mathbf{q}_{3}\cdot\mathbf{q}_{4})\left\langle \phi_{\mathbf{k}_{1}}\phi_{\mathbf{k}_{2}}\right\rangle _{\phi}\tilde{\theta}_{\mathbf{q}_{3}}\tilde{\theta}_{\mathbf{q}_{4}}\label{eq:L2}\\
L_{3}^{(1)}=4\times & L\int\frac{d^{d}k_{1}d^{d}k_{3}d^{d}q_{2}d^{d}q_{4}}{(2\pi)^{d}}\delta(\mathbf{k}_{1}+\mathbf{k}_{3}+\mathbf{q}_{2}+\mathbf{q}_{4})(\mathbf{k}_{1}\cdot\mathbf{q}_{2})(\mathbf{k}_{3}\cdot\mathbf{q}_{4})\left\langle \phi_{\mathbf{k}_{1}}\phi_{\mathbf{k}_{3}}\right\rangle _{\phi}\tilde{\theta}_{\mathbf{q}_{2}}\tilde{\theta}_{\mathbf{q}_{4}}.\label{eq:L3}
\end{align}
where integral over $\mathbf{q}$ is from $0$ to $\Lambda/b$ and
over $\mathbf{k}$ is from $\Lambda/b$ to $\Lambda$.

$L_{1}^{(1)}$ can be represented as a Feynman diagram in Table~\ref{tab:diagram}.
Here the solid lines indicate the low frequency modes. 
The dots represent dot product, \emph{i.e. }$\mathbf{q}_{1}\tilde{\theta}_{\mathbf{q}_{1}}$ is dotted with $\mathbf{q}_{2}\tilde{\theta}_{\mathbf{q}_{2}}$ and
$\mathbf{q}_{3}\tilde{\theta}_{\mathbf{q}_{3}}$ is dotted with $\mathbf{q}_{4}\tilde{\theta}_{4}$.
Finally the tiny dashed line between the two dots represents a momentum conservation $\delta(\mathbf{q}_{1}+\mathbf{q}_{2}+\mathbf{q}_{3}+\mathbf{q}_{4})$. 

Next, the $L_{2}^{(1)}$ term can be represented diagramatically as
in Table~\ref{tab:diagram}. Here the wavy lines represent high frequency
modes ($\mathbf{k}_{1}\phi_{\mathbf{k}_{1}}$ and $\mathbf{k}_{2}\phi_{\mathbf{k}_{2}}$).
These two wavy lines are connected to represent the correlation $\left\langle \phi_{\mathbf{k}_{1}}\phi_{\mathbf{k}_{2}}\right\rangle _{\phi}$
(which has another momentum conservation $\delta(\mathbf{k}_{1}+\mathbf{k}_{2})$
inside). Finally we can have $2$ different permutations of diagram
$L_{2}^{(1)}$ in Table~\ref{tab:diagram} (with wavy loop on the
left or on the right hand side) and thus we have a prefactor of $2$
in Eq.~(\ref{eq:L2}). We can now calculate $L_{2}^{(1)}$ term explicitly
\begin{align}
L_{2}^{(1)} = 2\times L\int\frac{d^{d}k_{1}d^{d}k_{2}d^{d}q_{3}d^{d}q_{4}}{(2\pi)^{d}}\delta(\mathbf{k}_{1}+\mathbf{k}_{2}+\mathbf{q}_{3}+\mathbf{q}_{4})(\mathbf{k}_{1}\cdot\mathbf{k}_{2})(\mathbf{q}_{3}\cdot\mathbf{q}_{4})\frac{\delta(\mathbf{k}_{1}+\mathbf{k}_{2})}{Kk_{1}^{2}+Ck_{1}^{4}}\tilde{\theta}_{\mathbf{q}_{3}}\tilde{\theta}_{\mathbf{q}_{4}}
\end{align}
where we have substituted $\left\langle \phi_{\mathbf{k}_{1}}\phi_{\mathbf{k}_{2}}\right\rangle _{\phi}=\frac{\delta(\mathbf{k}_{1}+\mathbf{k}_{2})}{Kk_{1}^{2}+Ck_{1}^{4}}$.
We next perform integral over $\mathbf{k}_{2}$ and then over $\mathbf{q}_{4}$
to eliminate the delta functions. Finally we obtain (after relabelling
$\mathbf{k}_{1}\rightarrow\mathbf{k}$ and $\mathbf{q}_{3}\rightarrow\mathbf{q}$):
\begin{equation}
L_{2}^{(1)}=2L\int_{0}^{\Lambda/b}d^{d}q\,q^{2}\left|\tilde{\theta}_{\mathbf{q}}\right|^{2}\int_{\Lambda/b}^{\Lambda}\frac{d^{d}k}{(2\pi)^{d}}\frac{k^{2}}{Kk^{2}+Ck^{4}}
\end{equation}

Similarly, the $L_{3}^{(1)}$ term can be represented diagramatically
as in Table~\ref{tab:diagram}. Here we have $4$ different permutations
and hence a prefactor of $4$ in Eq.~(\ref{eq:L3}). Again this integral
can be computed as:
\begin{align}
L_{3}^{(1)} & =4\times L\int\frac{d^{d}k_{1}d^{d}k_{3}d^{d}q_{2}d^{d}q_{4}}{(2\pi)^{d}}\delta(\mathbf{k}_{1}+\mathbf{k}_{3}+\mathbf{q}_{2}+\mathbf{q}_{4})(\mathbf{k}_{1}\cdot\mathbf{q}_{2})(\mathbf{k}_{3}\cdot\mathbf{q}_{4})\frac{\delta(\mathbf{k}_{1}+\mathbf{k}_{3})}{Kk_{1}^{2}+Ck_{1}^{4}}\tilde{\theta}_{\mathbf{q}_{2}}\tilde{\theta}_{\mathbf{q}_{4}}.
\end{align}
First we do integral over $\mathbf{k}_{3}$ then over $\mathbf{q}_{4}$
to eliminate the delta functions and then relabel $\mathbf{k}_{1}\rightarrow\mathbf{k}$
and $\mathbf{q}_{2}\rightarrow\mathbf{q}$ to obtain:
\begin{equation}
L_{3}^{(1)}=4\times L\int_{0}^{\Lambda/b}d^{d}q\,q_{\alpha}q_{\beta}\left|\tilde{\theta}_{\mathbf{q}}\right|^{2}\int_{\Lambda/b}^{\Lambda}\frac{d^{d}k}{(2\pi)^{d}}\frac{k_{\alpha}k_{\beta}}{Kk^{2}+Ck^{4}}.
\end{equation}
We observe that the integral over $\mathbf{k}$ is isotropic and symmetric
under swapping the indices $\alpha\leftrightarrow\beta$ and thus:
\begin{equation}
\int_{\Lambda/b}^{\Lambda}\frac{d^{d}k}{(2\pi)^{d}}\frac{k_{\alpha}k_{\beta}}{Kk^{2}+Ck^{4}}=A\delta_{\alpha\beta}
\end{equation}
for some constant $A$. Contracting the index $\alpha$ and $\beta$,
we can get $A$. Therefore
\begin{equation}
L_{3}^{(1)}=4L\int_{0}^{\Lambda/b}d^{d}q\,q^{2}\left|\tilde{\theta}_{\mathbf{q}}\right|^{2}\dfrac{1}{d}\int_{\Lambda/b}^{\Lambda}\frac{d^{d}k}{(2\pi)^{d}}\frac{1}{K+Ck^{2}}.
\end{equation}

Finally, the $\mathcal{O}(L)$ correction is
\begin{align}
\left\langle H_{I}\right\rangle _{\phi} & =L_{1}^{(1)}+L_{2}^{(1)}+L_{3}^{(1)}\\
 & =\frac{1}{2}\int_{0}^{\Lambda/b}d^{d}q\,q^{2}\left|\tilde{\theta}_{\mathbf{q}}\right|^{2}\left\{ 2L\left(2+\frac{4}{d}\right)\int_{\Lambda/b}^{\Lambda}\frac{d^{d}k}{(2\pi)^{d}}\frac{1}{K+Ck^{2}}\right\} \nonumber \\
 & +L\int_{0}^{\Lambda/b}\frac{d^{d}q_{1}\ldots d^{d}q_{4}}{(2\pi)^{d}}\delta(\mathbf{q}_{1}+\cdots\mathbf{q}_{4})(\mathbf{q}_{1}\cdot\mathbf{q}_{2})(\mathbf{q}_{3}\cdot\mathbf{q}_{4})\tilde{\theta}_{\mathbf{q}_{1}}\tilde{\theta}_{\mathbf{q}_{2}}\tilde{\theta}_{\mathbf{q}_{3}}\tilde{\theta}_{\mathbf{q}_{4}}
\end{align}

\subsection{$\mathcal{O}(L^{2})$ correction to $\tilde{H}[\tilde{\theta}_{\mathbf{q}}]$}

Next we calculate the $\mathcal{O}(L^{2})$ correction to the coarse-grained
Hamiltonian, \emph{i.e. }the second term in Eq.~(\ref{eq:H-coarse-grain}):
\begin{equation}
-\frac{1}{2}\left(\left\langle H_{I}^{2}\right\rangle _{\phi}-\left\langle H_{I}\right\rangle _{\phi}^{2}\right)=L_{4}^{(2)}+L_{5}^{(2)}+L_{6}^{(2)}
\end{equation}
The non-trivial terms are:
\begin{align}
L_{4}^{(2)} & =8\times\frac{-L^{2}}{2}\int\frac{d^{d}q_{1}d^{d}q_{2}d^{d}k_{3}d^{d}k_{4}d^{d}k_{5}d^{d}k_{6}d^{d}q_{7}d^{d}q_{8}}{(2\pi)^{2d}}\delta(\mathbf{q}_{1}+\mathbf{q}_{2}+\mathbf{k}_{3}+\mathbf{k}_{4})\delta(\mathbf{k}_{5}+\mathbf{k}_{6}+\mathbf{q}_{7}+\mathbf{q}_{8})\nonumber \\
 & (\mathbf{q}_{1}\cdot\mathbf{q}_{2})(\mathbf{k}_{3}\cdot\mathbf{k}_{4})(\mathbf{k}_{5}\cdot\mathbf{k}_{6})(\mathbf{q}_{7}\cdot\mathbf{q}_{8})\tilde{\theta}_{\mathbf{q}_{1}}\tilde{\theta}_{\mathbf{q}_{2}}\tilde{\theta}_{\mathbf{q}_{7}}\tilde{\theta}_{\mathbf{q}_{8}}\left\langle \phi_{\mathbf{k}_{3}}\phi_{\mathbf{k}_{5}}\right\rangle _{\phi}\left\langle \phi_{\mathbf{k}_{4}}\phi_{\mathbf{k}_{6}}\right\rangle _{\phi}\label{eq:L4}\\
L_{5}^{(2)} & =32\times\frac{-L^{2}}{2}\int\frac{d^{d}q_{1}d^{d}q_{2}d^{d}k_{3}d^{d}k_{4}d^{d}q_{5}d^{d}k_{6}d^{d}q_{7}d^{d}k_{8}}{(2\pi)^{2d}}\delta(\mathbf{q}_{1}+\mathbf{q}_{2}+\mathbf{k}_{3}+\mathbf{k}_{4})\delta(\mathbf{q}_{5}+\mathbf{k}_{6}+\mathbf{q}_{7}+\mathbf{k}_{8})\nonumber \\
 & (\mathbf{q}_{1}\cdot\mathbf{q}_{2})(\mathbf{k}_{3}\cdot\mathbf{k}_{4})(\mathbf{q}_{5}\cdot\mathbf{k}_{6})(\mathbf{q}_{7}\cdot\mathbf{k}_{8})\tilde{\theta}_{\mathbf{q}_{1}}\tilde{\theta}_{\mathbf{q}_{2}}\tilde{\theta}_{\mathbf{q}_{5}}\tilde{\theta}_{\mathbf{q}_{7}}\left\langle \phi_{\mathbf{k}_{3}}\phi_{\mathbf{k}_{6}}\right\rangle _{\phi}\left\langle \phi_{\mathbf{k}_{4}}\phi_{\mathbf{k}_{8}}\right\rangle _{\phi}\label{eq:L5}\\
L_{6}^{(2)} & =32\times\frac{-L^{2}}{2}\int\frac{d^{d}k_{1}d^{d}q_{2}d^{d}k_{3}d^{d}q_{4}d^{d}q_{5}d^{d}k_{6}d^{d}q_{7}d^{d}k_{8}}{(2\pi)^{2d}}\delta(\mathbf{k}_{1}+\mathbf{q}_{2}+\mathbf{k}_{3}+\mathbf{q}_{4})\delta(\mathbf{q}_{5}+\mathbf{k}_{6}+\mathbf{q}_{7}+\mathbf{k}_{8})\nonumber \\
 & (\mathbf{k}_{1}\cdot\mathbf{q}_{2})(\mathbf{k}_{3}\cdot\mathbf{q}_{4})(\mathbf{q}_{5}\cdot\mathbf{k}_{6})(\mathbf{q}_{7}\cdot\mathbf{k}_{8})\tilde{\theta}_{\mathbf{q}_{2}}\tilde{\theta}_{\mathbf{q}_{4}}\tilde{\theta}_{\mathbf{q}_{5}}\tilde{\theta}_{\mathbf{q}_{7}}\left\langle \phi_{\mathbf{k}_{1}}\phi_{\mathbf{k}_{6}}\right\rangle _{\phi}\left\langle \phi_{\mathbf{k}_{3}}\phi_{\mathbf{k}_{8}}\right\rangle _{\phi},\label{eq:L6}
\end{align}
where the integral over $\mathbf{q}$ is from $0$ to $\Lambda/b$
and the integral over $\mathbf{k}$ is from $\Lambda/b$ to $\Lambda$.
They are represented as Feynman diagrams in Table.~\ref{tab:diagram}.
The prefactors $8$, $32$, and $32$ represent the number of permutations
of these diagrams. Note that for every pair of two connected wavy
lines ($e.g.$ $\phi_{\mathbf{k}_{3}}$ and $\phi_{\mathbf{k}_{5}}$
in $L_{4}^{(2)}$) represents a single pair correlation $\left\langle \phi_{\mathbf{k}_{3}}\phi_{\mathbf{k}_{5}}\right\rangle _{\phi}$.
Feynman diagrams already take into account of Wick's theorem: $\left\langle \phi_{1}\phi_{2}\phi_{3}\phi_{4}\right\rangle _{\phi}=\left\langle \phi_{1}\phi_{2}\right\rangle _{\phi}\left\langle \phi_{3}\phi_{4}\right\rangle _{\phi}+\left\langle \phi_{1}\phi_{3}\right\rangle _{\phi}\left\langle \phi_{2}\phi_{4}\right\rangle _{\phi}+\left\langle \phi_{1}\phi_{4}\right\rangle _{\phi}\left\langle \phi_{2}\phi_{3}\right\rangle _{\phi}$
. The results over the integrals (\ref{eq:L4}-\ref{eq:L6}) are:
\begin{align}
L_{4}^{(2)} & =-4L^{2}\int_{0}^{\Lambda/b}\frac{d^{d}q_{1}\ldots d^{d}q_{4}}{(2\pi)^{d}}\delta(\mathbf{q}_{1}+\cdots\mathbf{q}_{4})(\mathbf{q}_{1}\cdot\mathbf{q}_{2})(\mathbf{q}_{3}\cdot\mathbf{q}_{4})\tilde{\theta}_{\mathbf{q}_{1}}\tilde{\theta}_{\mathbf{q}_{2}}\tilde{\theta}_{\mathbf{q}_{3}}\tilde{\theta}_{\mathbf{q}_{4}}\left(\int_{\Lambda/b}^{\Lambda}\frac{d^{d}k}{(2\pi)^{d}}\frac{1}{(K+Ck^{2})^{2}}\right)\label{eq:L4-2}\\
L_{5}^{(2)} & =-\frac{16}{d}L^{2}\int_{0}^{\Lambda/b}\frac{d^{d}q_{1}\ldots d^{d}q_{4}}{(2\pi)^{d}}\delta(\mathbf{q}_{1}+\cdots\mathbf{q}_{4})(\mathbf{q}_{1}\cdot\mathbf{q}_{2})(\mathbf{q}_{3}\cdot\mathbf{q}_{4})\tilde{\theta}_{\mathbf{q}_{1}}\tilde{\theta}_{\mathbf{q}_{2}}\tilde{\theta}_{\mathbf{q}_{3}}\tilde{\theta}_{\mathbf{q}_{4}}\left(\int_{\Lambda/b}^{\Lambda}\frac{d^{d}k}{(2\pi)^{d}}\frac{1}{(K+Ck^{2})^{2}}\right)\label{eq:L5-2}\\
L_{6}^{(2)} & =-\frac{48}{d(d+2)}L^{2}\int_{0}^{\Lambda/b}\frac{d^{d}q_{1}\ldots d^{d}q_{4}}{(2\pi)^{d}}\delta(\mathbf{q}_{1}+\cdots\mathbf{q}_{4})(\mathbf{q}_{1}\cdot\mathbf{q}_{2})(\mathbf{q}_{3}\cdot\mathbf{q}_{4})\tilde{\theta}_{\mathbf{q}_{1}}\tilde{\theta}_{\mathbf{q}_{2}}\tilde{\theta}_{\mathbf{q}_{3}}\tilde{\theta}_{\mathbf{q}_{4}}\left(\int_{\Lambda/b}^{\Lambda}\frac{d^{d}k}{(2\pi)^{d}}\frac{1}{(K+Ck^{2})^{2}}\right)\label{eq:L6-2}
\end{align}

Here we will only show the result of (\ref{eq:L6-2}) since the results
of (\ref{eq:L4-2}-\ref{eq:L5-2}) are similar. From (\ref{eq:L6}),
we have:
\begin{align}
L_{6}^{(2)} & =-16L^{2}\int\frac{d^{d}k_{1}d^{d}q_{2}d^{d}k_{3}d^{d}q_{4}d^{d}q_{5}d^{d}k_{6}d^{d}q_{7}d^{d}k_{8}}{(2\pi)^{2d}}\delta(\mathbf{k}_{1}+\mathbf{q}_{2}+\mathbf{k}_{3}+\mathbf{q}_{4})\delta(\mathbf{q}_{5}+\mathbf{k}_{6}+\mathbf{q}_{7}+\mathbf{k}_{8})\nonumber \\
 & (\mathbf{k}_{1}\cdot\mathbf{q}_{2})(\mathbf{k}_{3}\cdot\mathbf{q}_{4})(\mathbf{q}_{5}\cdot\mathbf{k}_{6})(\mathbf{q}_{7}\cdot\mathbf{k}_{8})\tilde{\theta}_{\mathbf{q}_{2}}\tilde{\theta}_{\mathbf{q}_{4}}\tilde{\theta}_{\mathbf{q}_{5}}\tilde{\theta}_{\mathbf{q}_{7}}\frac{\delta(\mathbf{k}_{1}+\mathbf{k}_{6})}{Kk_{1}^{2}+Ck_{1}^{4}}\frac{\delta(\mathbf{k}_{3}+\mathbf{k}_{8})}{Kk_{3}^{2}+Ck_{3}^{4}}
\end{align}
Next, we integrate over $\mathbf{k}_{6}$ and $\mathbf{k}_{8}$ to
eliminate $\delta(\mathbf{k}_{1}+\mathbf{k}_{6})$ and $\delta(\mathbf{k}_{3}+\mathbf{k}_{8})$
and then over $\mathbf{k}_{3}$ to eliminate another delta function.
The result is (after relabelling $\mathbf{k}_{1}\rightarrow\mathbf{k}$):
\begin{align}
L_{6}^{(2)} & =-16L^{2}\int\frac{d^{d}q_{2}d^{d}q_{4}d^{d}q_{5}d^{d}q_{7}}{(2\pi)^{d}}\delta(\mathbf{q}_{2}+\mathbf{q}_{4}+\mathbf{q}_{5}+\mathbf{q}_{7})\tilde{\theta}_{\mathbf{q}_{2}}\tilde{\theta}_{\mathbf{q}_{4}}\tilde{\theta}_{\mathbf{q}_{5}}\tilde{\theta}_{\mathbf{q}_{7}}\nonumber \\
 & \int\frac{d^{d}k}{(2\pi)^{d}}(\mathbf{q}_{2}\cdot\mathbf{k})(\mathbf{q}_{5}\cdot\mathbf{k})\frac{\mathbf{q}_{4}\cdot(\mathbf{k}-\mathbf{q}_{5}-\mathbf{q}_{7})}{Kk^{2}+Ck^{4}}\frac{\mathbf{q}_{7}\cdot(\mathbf{k}-\mathbf{q}_{5}-\mathbf{q}_{7})}{K\left|\mathbf{k}-\mathbf{q}_{5}-\mathbf{q}_{7}\right|^{2}+C\left|\mathbf{k}-\mathbf{q}_{5}-\mathbf{q}_{7}\right|^{4}}
\end{align}
Then we can assume $\left|\mathbf{k}\right|\gg\left|\mathbf{q}_{5}+\mathbf{q}_{7}\right|$
and thus we obtain:
\begin{align}
L_{6}^{(2)} & =-16L^{2}\int\frac{d^{d}q_{2}d^{d}q_{4}d^{d}q_{5}d^{d}q_{7}}{(2\pi)^{d}}\delta(\mathbf{q}_{2}+\mathbf{q}_{4}+\mathbf{q}_{5}+\mathbf{q}_{7})q_{2\alpha}q_{4\beta}q_{5\gamma}q_{7\delta}\tilde{\theta}_{\mathbf{q}_{2}}\tilde{\theta}_{\mathbf{q}_{4}}\tilde{\theta}_{\mathbf{q}_{5}}\tilde{\theta}_{\mathbf{q}_{7}}\int\frac{d^{d}k}{(2\pi)^{d}}\frac{k_{\alpha}k_{\beta}k_{\gamma}k_{\delta}}{\left(Kk^{2}+Ck^{4}\right)^{2}}
\end{align}
Next, we observe that the $\mathbf{k}$-integral is isotropic
and symmetric under swapping any indices $\alpha\leftrightarrow\beta$,
$\alpha\leftrightarrow\gamma$, \emph{etc. }Therefore the $\mathbf{k}$-integral
can be written as:
\begin{equation}
\int\frac{d^{d}k}{(2\pi)^{d}}\frac{k_{\alpha}k_{\beta}k_{\gamma}k_{\delta}}{\left(Kk^{2}+Ck^{4}\right)^{2}}=B(\delta_{\alpha\beta}\delta_{\gamma\delta}+\delta_{\alpha\delta}\delta_{\beta\gamma}+\delta_{\alpha\gamma}\delta_{\beta\delta}),
\end{equation}
for some constant $B$. Contracting all the indices, we can obtain
this constant $B$ and the result is (\ref{eq:L6-2}).

Therefore the coarse-grained Hamiltonian Eq.~(\ref{eq:H-coarse-grain}),
after averaging out the high-frequency modes, is:
\begin{align}
\tilde{H}[\tilde{\theta}_{\mathbf{q}}] & =H_{0}[\tilde{\theta}_{\mathbf{q}}]+L_{1}^{(1)}+L_{2}^{(1)}+L_{3}^{(1)}+L_{4}^{(2)}+L_{5}^{(2)}+L_{6}^{(2)}\\
 & =\frac{1}{2}\int_{0}^{\Lambda/b}d^{d}q\left(\tilde{K}q^{2}+Cq^{4}\right)\left|\tilde{\theta}_{\mathbf{q}}\right|^{2}\nonumber \\
 & +\tilde{L}\int_{0}^{\Lambda/b}\frac{d^{d}q_{1}d^{d}q_{2}d^{d}q_{3}}{(2\pi)^{d}}\delta(\mathbf{q}_{1}+\mathbf{q}_{2}+\mathbf{q}_{3})\left[\mathbf{q}_{1}\cdot\mathbf{q}_{2}\right]\left[\mathbf{q}_{3}\cdot(-\mathbf{q}_{1}-\mathbf{q}_{2}-\mathbf{q}_{3})\right]\tilde{\theta}_{\mathbf{q}_{1}}\tilde{\theta}_{\mathbf{q}_{2}}\tilde{\theta}_{\mathbf{q}_{3}}\tilde{\theta}_{-\mathbf{q}_{1}-\mathbf{q}_{2}-\mathbf{q}_{3}}\label{eq:H-coarse-grain-2}
\end{align}
where
\begin{align}
\tilde{K} & =K+4L\left(1+\frac{2}{d}\right)\int_{\Lambda/b}^{\Lambda}\frac{d^{d}k}{(2\pi)^{d}}\frac{1}{K+Ck^{2}}\\
\tilde{L} & =L-4L^{2}\left(1+\frac{4}{d}+\frac{12}{d(d+2)}\right)\int_{\Lambda/b}^{\Lambda}\frac{d^{d}k}{(2\pi)^{d}}\frac{1}{(K+Ck^{2})^{2}}
\end{align}

\subsection{Rescaling $\mathbf{q}\rightarrow b\mathbf{q}$ and $\tilde{\theta}\rightarrow\tilde{\theta}/z$}

\begin{figure}
\begin{centering}
\includegraphics[width=0.8\columnwidth]{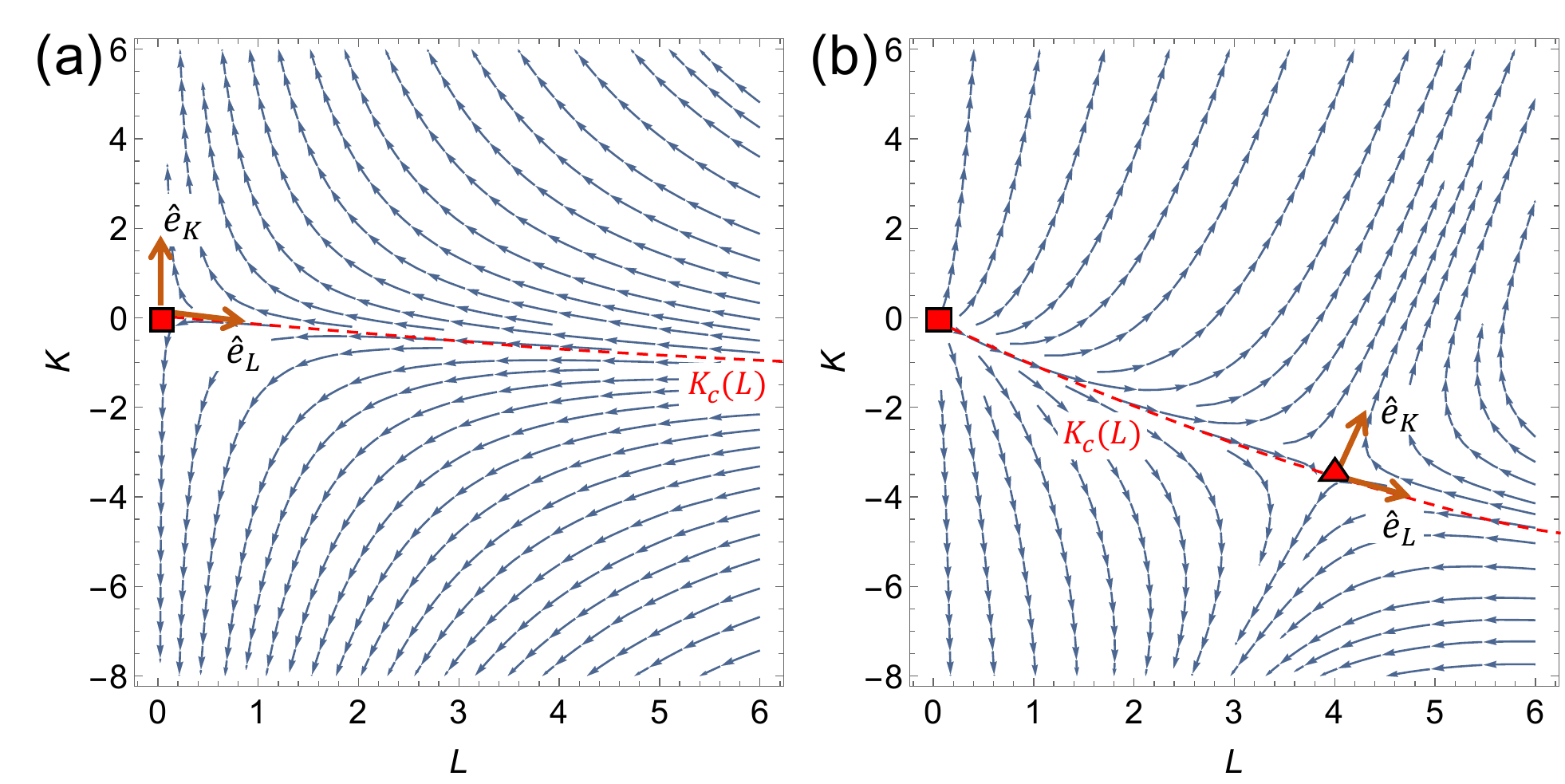}
\par\end{centering}
\caption{
Renormalization group (RG) flow for the parameters $K=\frac{K_{1}+K_{3}}{2}$ and $L$ for $d\ge4$ (a) and $d<4$ $(b)$. 
Square and triangle are the fixed points $(0,0)$ and $(K^*,L^*)$ respectively.
\label{fig:RG-flow}}
\end{figure}

The second and final step in RG  (see Fig.~\ref{fig:RG}) is to rescale
$\mathbf{q}$ and $\tilde\theta$. In particular, we define $\mathbf{q}'=b\mathbf{q}$
and $\theta'=\tilde{\theta}/z$ where $b>1$ is the rescaling factor.
Eq.~(\ref{eq:H-coarse-grain-2}) then becomes:
\begin{align}
H'[\theta'_{\mathbf{q}'}] & =\frac{1}{2}\int_{0}^{\Lambda}d^{d}q'\left(K'q'^{2}+C'q'^{4}\right)\left|\theta'_{\mathbf{q}'}\right|^{2}\nonumber \\
 & +L'\int_{0}^{\Lambda}\frac{d^{d}q_{1}'d^{d}q_{2}'d^{d}q_{3}'}{(2\pi)^{d}}\delta(\mathbf{q}_{1}'+\mathbf{q}_{2}'+\mathbf{q}_{3}')\left[\mathbf{q}_{1}'\cdot\mathbf{q}_{2}'\right]\left[\mathbf{q}_{3}'\cdot(-\mathbf{q}_{1}'-\mathbf{q}_{2}'-\mathbf{q}_{3}')\right]\theta'_{\mathbf{q}_{1}'}\theta'_{\mathbf{q}_{2}'}\theta'_{\mathbf{q}_{3}'}\theta'_{-\mathbf{q}_{1}'-\mathbf{q}_{2}'-\mathbf{q}_{3}'}.\label{eq:rescaled-Hamiltonian}
\end{align}
where 
\begin{align}
K' & =b^{-d-2}z^{2}\left[K+4L\left(1+\frac{2}{d}\right)\int_{\Lambda/b}^{\Lambda}\frac{d^{d}k}{(2\pi)^{d}}\frac{1}{K+Ck^{2}}\right]\label{eq:RG1}\\
C' & =b^{-d-4}z^{2}C\label{eq:RG2}\\
L' & =b^{-3d-4}z^{4}\left[L-4L^{2}\left(1+\frac{4}{d}+\frac{12}{d(d+2)}\right)\int_{\Lambda/b}^{\Lambda}\frac{d^{d}k}{(2\pi)^{d}}\frac{1}{(K+Ck^{2})^{2}}\right].\label{eq:RG3}
\end{align}
Therefore at the end of RG, we end up with the same form of Hamiltonian but with renormalized coefficients $K'$, $C'$, and $L'$. 
Now we need to fix $z$ by choosing the coefficient $C$ to be invariant under RG. 
The reason is because we are more interested in
how the coefficients $K$ (which is the control parameter in our model)
and $L$ change under RG. If $L$ becomes smaller when we coarse-grain
and rescale, the higher order term in $H$ is shown to be irrelevant.
Thus we fix $C'=C$ to obtain $z=b^{(d+4)/2}$. Next we rewrite $b$
as $b=1+\delta\ell$ where $\delta\ell$ is small and positive. The
integral in Eq.~(\ref{eq:RG1}) then becomes:
\begin{equation}
\int_{\Lambda-\Lambda\delta\ell}^{\Lambda}\frac{1}{K+Ck^{2}}  \frac{\Omega_d}{(2\pi)^d} k^{d-1}dk \simeq \frac{1}{K+C\Lambda^{2}} \frac{\Omega_d}{(2\pi)^d} \Lambda^{d-1}\Lambda\delta\ell
\end{equation}
where $\Omega_d$ is the solid angle in a $d$-dimensional sphere. 
Similarly, we can compute the integral in Eq.~(\ref{eq:RG3}) over a thin shell of radius $\Lambda$ and thickness $\Lambda\delta\ell$. 
We then obtain the RG flow for the coefficients $K$ and $L$:
\begin{align}
\frac{dK}{d\ell} & =2K + 4\left(1+\frac{2}{d}\right) \frac{\Omega_d}{(2\pi)^d} \frac{\Lambda^{d}}{K+C\Lambda^{2}} L \label{eq:K}\\
\frac{dL}{d\ell} & =(4-d)L - 4\left(1+\frac{4}{d}+\frac{12}{d(d+2)}\right) \frac{\Omega_d}{(2\pi)^d} \frac{\Lambda^{d}}{(K+C\Lambda^{2})^{2}} L^2 .\label{eq:L}
\end{align}
The RG flow for the coefficients $K$ and $L$ are plotted in Fig.~\ref{fig:RG-flow}.

For spatial dimension $d\ge4$, (\ref{eq:K}-\ref{eq:L}) have one
non-trivial fixed point at $(K=0,L=0)$, which is unstable along $\hat{e}_{K}$-direction
and stable along $\hat{e}_{L}$-direction (see Fig.~\ref{fig:RG-flow}(a)).
This means if we are above the critical line $K_{c}(L)$ (dashed red
line in Fig.~\ref{fig:RG-flow}(a)), RG flow will take us to $K\rightarrow+\infty$
(uniform nematic phase) and if we are below $K_{c}(L)$, RG flow will
take us to $K\rightarrow-\infty$ ($SB_{\infty}$ phase). Linearizing
(\ref{eq:K}-\ref{eq:L}) around this fixed point, we obtain:
\begin{equation}
\frac{d}{d\ell}\left(\begin{array}{c}
\delta K\\
\delta L
\end{array}\right)=\left(\begin{array}{cc}
2 & 4\left(1+\frac{2}{d}\right) \frac{\Omega_d}{(2\pi)^d} \frac{\Lambda^{d}}{C\Lambda^{2}}\\
0 & 4-d
\end{array}\right)\left(\begin{array}{c}
\delta K\\
\delta L
\end{array}\right)
\end{equation}
with eigenvalue $\lambda_{K}=2$ associated with eigendirection $\hat{e}_{K}$
and $\lambda_{L}=4-d\le0$ associated with eigendirection $\hat{e}_{L}$
(see Fig.~\ref{fig:RG-flow}(a)). Thus close to the fixed point,
$\delta K$ diverges as $\left|\delta K\right|\sim e^{\lambda_{K}\ell}$.
Now from Fig.~\ref{fig:RG}, under an infinitesimal RG, the correlation
length $\xi$ is mapped to $\xi'$, which is given by:
\begin{align}
\xi' & =\frac{\xi}{(1+\delta\ell)}\simeq\xi-\delta\ell\xi\quad\Rightarrow\quad\frac{d\xi}{d\ell}=-\xi\quad\Rightarrow\quad\xi\sim e^{-\ell}
\end{align}
Substituting $\ell\sim\ln\left|\delta K\right|/\lambda_{K}$, we obtain:
\begin{equation}
\xi\sim\left|\delta K\right|^{-\frac{1}{\lambda_{K}}}
\end{equation}
Therefore for $d\ge4$ we get the mean field critical exponent $\nu=1/\lambda_{K}=1/2$
as we expect. However, RG calculation also tells us that the critical
point $K_{c}(L)$ is not zero (as shown by the mean field phase diagram
in the main text) but shifted by $L$.

Now for spatial dimension $d<4$, the fixed point at $(0,0)$ becomes
unstable and a new non-trivial fixed point appears at $(K^{*}=-\frac{3}{10}C\Lambda^{2}\epsilon,L^{*}=\frac{4\pi^{2}}{5}C^{2}\epsilon)$,
where $\epsilon=4-d>0$ (see Fig.~\ref{fig:RG-flow}(b)). Linearizing
around $(K^{*},L^{*})$ and expanding for small $\epsilon$, we get:
\begin{equation}
\frac{d}{d\ell}\left(\begin{array}{c}
\delta K\\
\delta L
\end{array}\right)=\left(\begin{array}{cc}
2-\frac{3}{5}\epsilon+\mathcal{O}(\epsilon^{2}) & \frac{3\Lambda^{2}}{4\pi^{2}C}+\mathcal{O}(\epsilon)\\
\mathcal{O}(\epsilon^{2}) & -\epsilon+\mathcal{O}(\epsilon^{2})
\end{array}\right)\left(\begin{array}{c}
\delta K\\
\delta L
\end{array}\right).
\end{equation}
Thus we identify the eigenvalue $\lambda_{K}=2-\frac{3}{4}\epsilon+\mathcal{O}(\epsilon^{2})$
associated with the eigendirection $\hat{e}_{K}$ (see Fig.~\ref{fig:RG-flow}(b)).
Therefore the critical exponent for the correlation length $\xi$
is:
\begin{equation}
\nu=\frac{1}{\lambda_{K}}=\frac{1}{2}+\frac{3}{20}\epsilon+\mathcal{O}(\epsilon^{2}).
\end{equation}

\subsection{Comparison with ferromagnet}
In comparison, for unconstrained para-ferromagnetic transition, the Hamiltonian is given by
\begin{equation}
H[\mathbf{m}]=\int d^dr\left\{ \frac{K}{2}\left|\mathbf{m}\right|^2 + \frac{C}{2}\left|\nabla\mathbf{m}\right|^2 + L \left|\mathbf{m}\right|^4 \right\},
\end{equation}
and the RG flow equations for $K$ and $L$ are:
\begin{eqnarray}
\frac{dK}{d\ell} &=& 2K + 4(n+2)\frac{\Lambda^d}{K+C\Lambda^2}\frac{\Omega_d}{(2\pi)^d}L \\
\frac{dL}{d\ell} &=& (4-d)L - 4(n+8)\frac{\Lambda^d}{\left(K+C\Lambda^2\right)^2}\frac{\Omega_d}{(2\pi)^d}L^2,
\end{eqnarray}
where $n$ is the dimension of $\mathbf{m}$ and $d$ is the spatial dimension.


\begin{thebibliography}{10}
\bibitem{curl}
Note that in $d$-dimension, $\theta$ is a function on the manifold $\mathbb{R}^d$.
$\nabla\theta$ is the exterior derivative of $\theta$, which is a $1$-form.
``$\nabla\times\nabla\theta$'' is taking another exterior derivative, which is zero.
For example in $d=2$, $\nabla\times\mathbf{m}=\epsilon_{\alpha\beta}\partial_\alpha m_\beta$ is a scalar.
\end{thebibliography}
\end{document}